\documentclass[preprints,review,accept,pdftex,moreauthors]{Definitions/mdpi}

\firstpage{1} 
\pubvolume{10}
\issuenum{10}
\articlenumber{1958}
\pubyear{2022}
\copyrightyear{2022}
\externaleditor{Academic Editor: { Emma Previato and Vladimir Balan}}
\datereceived{11 April 2022} 
\dateaccepted{3 June 2022} 
\datepublished{7 June 2022} 
\hreflink{https://doi.org/10.3390/math10121958}

\usepackage{subcaption}

\usepackage{amsmath,amsfonts,bbm,dsfont,mathrsfs}
\allowdisplaybreaks
\newcommand{\half}{\frac{1}{2}}
\newcommand{\dd}{{\mathrm{d}}}
\newcommand{\tr}{\tilde{r}}
\newcommand{\tlt}{\tilde{t}}
\newcommand{\ta}{\tilde{a}}
\newcommand{\tL}{\tilde{L}}
\newcommand{\tR}{\tilde{R}}
\newcommand{\tXi}{\tilde{\Xi}}
\newcommand{\tA}{\tilde{A}}
\newcommand{\tB}{\tilde{B}}
\newcommand{\tTh}{\tilde{\Theta}}
\newcommand{\tK}{\tilde{K}}
\newcommand{\tN}{\tilde{N_g}}
\newcommand{\te}{\tilde{e}}
\newcommand{\tg}{\tilde{g}}
\newcommand{\tv}{\tilde{v}}

\Title{Hyperelliptic Functions and Motion in General Relativity}

\TitleCitation{Hyperelliptic Functions and Motion in General Relativity}

\Author{Saskia Grunau $^{1,\dagger}$ and Jutta Kunz $^{2,\dagger}$}

\AuthorNames{Saskia Grunau and Jutta Kunz}

\AuthorCitation{Grunau, S.; Kunz, J.}

\address{%
$^{1}$ \quad saskia.grunau@uni-oldenburg.de\\
$^{2}$ \quad jutta.kunz@uni-oldenburg.de}

\firstnote{Institut für Physik, Universität Oldenburg, D-26111 Oldenburg, Germany.}

\abstract{Analysis of black hole spacetimes requires study of the motion of particles and light in these spacetimes.
Here exact solutions of the geodesic equations are the means of choice.
Numerous interesting black hole spacetimes have been analyzed in terms of elliptic functions.
However, the presence of a cosmological constant, higher dimensions or alternative gravity theories often necessitate an analysis in terms of hyperelliptic functions. 
Here we review the method and current status for solving the geodesic equations for the general hyperelliptic case, illustrating it with a set of examples of genus $g=2$:
higher dimensional Schwarzschild black holes, rotating dyonic $U(1)^2$ black holes, and black rings.
}



\begin{document}
\section{Introduction}

More than one hundred years ago, Einstein proposed his theory of general relativity, based on the revolutionary idea, that gravity is encoded in the geometric properties of space and time.
General relativity is very well supported by experiments and has many applications such as the global positioning system GPS~\cite{Ashby:2003, Will:2018bme}.
One of the predictions of general relativity is the existence of black holes and their formation in the collapse of very massive stars after exhaustion of their nuclear fuel~\cite{Oppenheimer:1939ue,Penrose:1964wq,Penrose:1969pc, Penrose:1969pc-2}.
Black holes possess an event horizon, and~thus a boundary beyond which no communication with the exterior is possible.
By now there is strong observational evidence not only for stellar black holes, but~also for supermassive black holes at the core of galaxies~\cite{Webster:1972bsw,Bolton:1972bun,Kormendy:1995er,Eckart:1996zz,Ghez:1998ph,Celotti:1999tg,Ferrarese:2004qr,LIGOScientific:2016aoc,EventHorizonTelescope:2019dse}.  

A powerful tool to study black holes is the analysis of their geodesics. 
The motion of particles and light around a black hole provides valuable information about the spacetime.
\textls[-15]{In particular, analytic solutions of the geodesic equations can be used to calculate observables with high accuracy, to~be compared with observations in order to test theories and~models.}

For many well-known spacetimes like Schwarzschild and Kerr the equations of motion are of elliptic type (see, e.g.,~\cite{Hagihara:1931,Carter:1968rr,Bardeen:1973,Chandrasekhar:1985kt,Perlick:2004tq,Kraniotis:2007zz,Kagramanova:2010bk,Grunau:2010gd,Hackmann:2010ir,Kraniotis:2010gx,Kagramanova:2012hw,Hackmann:2013pva,Diemer:2013hgn,Diemer:2013fza,Grunau:2013oca,Grenzebach:2014fha,Diemer:2014lba,Flathmann:2015xia,Kraniotis:2014paa,Kraniotis:2015kfd,Paranjape:2016oly,Grunau:2017uzf,Eickhoff:2018msh,Willenborg:2018zsv,Drawer:2020mpw} and references therein).
These can easily be solved analytically in terms of the elliptic and therefore doubly-periodic Weierstraß $\wp$-function. 
However, when additional parameters (like the cosmological constant), higher dimensions or alternative theories of gravity are considered, the~equations of motion become often more complicated and one encounters hyperelliptic integrals~\cite{Baker:1995,Kraniotis:2003ig,Kraniotis:2004cz,Kraniotis:2005zm}.

\textls[-25]{Inspired by the work on the double pendulum by Enolski~et~al.~\cite{Enolski:2003}, exact solutions of the hyperelliptic equations of motion of genus two ($g=2$) arising in Schwarzschild-(anti-) de Sitter spacetime were obtained by Hackmann and L\"ammerzahl~\cite{Hackmann:2008zza,Hackmann:2008zz}.
Subsequently exact solutions of the $g=2$ geodesic equations were obtained for spherically and axially symmetric black holes in four and higher dimensions~\cite{Hackmann:2008tu,Hackmann:2009nh,Hackmann:2010zz}.}
The inversion of hyperelliptic integrals was then generalized by Enolski~et~al.~\cite{Enolski:2010if} to obtain solutions of the geodesic equations also for higher genus, $g>2$.
First applications included a nine-dimensional spacetime with cosmological constant and charge~\cite{Enolski:2010if} and special cases of a Hořava–Lifshitz spacetime~\cite{Enolski:2011id}. 

\section{Geodesic Motion around Black~Holes}
\label{sec:GeoMotion}

Black holes represent some of the most intriguing objects of the universe, 
making their study both observationally and mathematically highly interesting. 
When the black hole spacetime is obtained as a solution of the gravitational field equations, the~exploration of the properties of this spacetime relies to a large extent on the analysis of the motion of particles and light in this spacetime.
For neutral point particles this motion is described by the geodesic equation
\begin{equation}
0 = \frac{d^2 x^\mu}{d\tau^2} + \left\{\begin{smallmatrix}\mu \\ \rho \sigma\end{smallmatrix}\right\} \frac{dx^\rho}{d\tau} \frac{dx^\sigma}{d\tau} \ ,
\label{geodesic}
\end{equation}
where $x^\mu$ are the coordinates, $\tau$ is an affine parameter along the orbit, and~the symbol $\left\{\begin{smallmatrix}\mu \\ \rho \sigma\end{smallmatrix}\right\}$ denotes the connection coefficients (and the Einstein summation convention is employed).
Given a spacetime described by the invariant square of the infinitesimal line element
\begin{equation}
d s^2 = g_{\mu\nu} dx^\mu dx^\nu
\end{equation}
with metric coefficients $g_{\mu\nu}$, the~connection coefficients are known functions of the coordinates.
The geodesic Equation~(\ref{geodesic}) forms a set of coupled differential equations, whose solution describes the sought after orbital motion in the spacetime.
However, it is often of advantage to apply the Hamilton--Jacobi formalism to obtain an equivalent set of equations, starting from the Hamilton--Jacobi equation
\begin{equation}
2\frac{\partial S}{\partial\tau} = g^{ \mu\nu}\frac{\partial S}{\partial x^\mu}\frac{\partial S}{\partial x^\nu} 
\label{HJ}
\end{equation}
with Hamilton's principal function $S$, and~to exploit the symmetries of the spacetime to obtain a separation of~variables.

Black holes in four dimensions, i.e.,~one time and three spatial dimensions, are certainly of highest interest from an astrophysical point of view. 
In this case separability requires four constants of motion.
The first constant of motion $\delta$ is associated with the square of the particle momentum, $\delta = p_\mu p^\mu = -m^2$.
Setting the particle mass to $m=1$ yields $\delta=-1$, whereas for light $\delta=0$.
The simplest black hole solutions of General Relativity are the static spherically symmetric Schwarzschild black hole and the stationary axially symmetric (rotating) Kerr black hole.
Clearly, the~symmetries of these spacetimes correspond to the existence of Killing vectors associated with two constants of motion.
In particular, stationarity leads to conservation of energy $E$ and axial symmetry to conservation of angular momentum $L$.
However, the~separability of the equations of motion of the Kerr spacetime involves still another constant of motion, the~so called Carter constant~\cite{Carter:1968rr}, that derives from the existence of a Killing tensor~\cite{Walker:1970un}.
In Boyer--Lindquist coordinates $t,r,\theta,\phi$ the geodesic equations can then be solved after employing the separation ansatz
\begin{equation}
S= \frac{1}{2} \delta \tau-{E} t+L\phi+S_{r}(r)+S_{\theta }(\theta)  \ ,
\label{ansatz}
\end{equation}
leading to elliptic integrals, just as for the case of the static Schwarzschild black~hole.

From a theoretical point of view it is very interesting to consider General Relativity also in higher dimensions, since additional dimensions are present in various theories like Kaluza--Klein theory or string theory.
Moreover, it allows to discern genuine properties of black holes in General Relativity and properties only present in four dimensions.
Higher dimensional generalizations of the static spherically symmetric Schwarzschild black hole were obtained by Tangherlini~\cite{Tangherlini:1963bw}.
The high symmetry of these solutions implies separability of the equations of motion, but~depending on the spacetime dimension $D$ hyperelliptic integrals arise, starting with $D=6$ \cite{Hackmann:2008tu}.
The generalizations of the rotating Kerr black hole to higher dimensions was accomplished by Myers and Perry~\cite{Myers:1986un}.
Myers--Perry black holes in $D$ dimensions are characterized by $N=\lfloor \frac{D-1}{2} \rfloor $ independent angular momenta, associated with rotation in $N$ independent planes.
Thus 5-dimensional Myers--Perry black holes possess two conserved angular momenta, showing the need for another conserved quantity for separability.
Analogous to the Kerr case separability is guaranteed based on the presence of a Killing tensor~\cite{Frolov:2002xf,Frolov:2003en}.
Analysis of the geodesic equations leads again to elliptic integrals~\cite{Kagramanova:2012hw,Diemer:2014lba}.
Proof of separability of the geodesic equations in still higher dimensions is based on the existence of Killing-Yano tensors~\cite{Page:2006ka,Kubiznak:2006kt}. 
In these spacetimes hyperelliptic integrals arise unless substitutions can be made to reduce the order of the polynomial $P_d$ in the geodesic equations.
As discussed in more detail below, hyperelliptic integrals arise also in numerous further black hole spacetimes, for~instance, when a cosmological constant, electromagnetic fields, etc.~are~present.

\section{Hyperelliptic~Integrals}
\label{sec:hyperell}
When the equations of motion are of hyperelliptic type they take the form
\begin{equation}
	\left( y^n \frac{\dd y}{\dd x} \right)^2 = P_d(y)
	\label{eqn:differentialeq}
\end{equation}
with the initial values $x_\mathrm{in}$ and $y_\mathrm{in}$, and~have to be solved for a function $y(x)$. $P_d(y)$ is a polynomial of order $d\geq 5$. The~number $n$ can take the values $n=1 \ldots g-1$ where $g=\lfloor \frac{d-1}{2} \rfloor$.

Equation \eqref{eqn:differentialeq} leads to a hyperelliptic integral of the first kind
\begin{equation}
	x-x_\mathrm{in} = \int_{y_\mathrm{in}}^{y}\! \frac{{y'}^n \ \dd y'}{\sqrt{ P_d(y')}}
	\label{eqn:hyper-firstkind}
\end{equation}
which has to be inverted to find $y(x)$. The~solution has been discussed in~\cite{Hackmann:2008tu,Hackmann:2008zza,Hackmann:2008zz,Enolski:2010if,Enolski:2011id}.  
The inversion of the integral should not depend on the integration path and thus for a closed integration path with the integral $\omega = \int\!\frac{{y}^n \ \dd y}{\sqrt{ P_d(y)}}$ the following must apply: $y(x-\omega)=y(x)$. Therefore $y(x)$ is periodic.
For each $y$ the square root in the integrand can take two different signs. It is therefore not clearly defined in the complex plane.
To resolve this issue, a~Riemann surface can be constructed, where $y\mapsto \sqrt{P_d(y)}$ is a single valued function.
The two signs of $\sqrt{P_d(y)}$ correspond to two copies of the Riemann sphere. Both spheres are cut between pairs of the zeros $e_i$ with $i=1 \ldots d$ of the polynomial $P_d(y)$. Here the zeros are called branch points. The~branch cuts between the branch points should not be right next to each other, i.e.,~there should not be two cuts starting at the same branch point.
If $d$ is odd, the~branch point $e_{d+1}$ is placed at infinity. 
At the branch points $\sqrt{P_d(y)}$ is the same at both spheres and therefore they can be identified here. Along the branch cuts the two spheres can be put together to get the Riemann surface, which {now has} holes. The~number of holes is described by the genus $g=\lfloor \frac{d-1}{2} \rfloor $. For~example in the case $g=2$, the~Riemann surface is a double torus with two~holes.

Let $P_d(y)=\lambda_{2g+1}y^{2g+1}+\lambda_{2g} y^{2g} + \ldots + \lambda_0 $ be the polynomial with one branch point at infinity. For~the problems considered in this review, the~polynomial is often considered in the canonical form, which is defined by setting the first coefficient $\lambda_{2g+1}=4$. The~factor 4 is chosen so that the canonical form looks like the Weierstraß form of the polynomial in the corresponding elliptic problem~\cite{Enolski:2010if}.

Then a basis of holomorphic differentials of the first kind $\dd u_i$ and meromorphic differentials of the second kind $\dd r_i$ with $i=1\ldots g$ is defined as
\begin{equation}
	\dd u_i=\frac{y^{i-1}\dd y}{\sqrt{P_d(y)}} \, , \quad 
	\dd r_i=\sum _{k=i}^{d-i}(k+1-i) \lambda _{k+1+i} \frac{y^{k}}{4\sqrt{P_d(y)}} \dd y \, .
	\label{eqn:hyper-differentials}
\end{equation}

A differential $\frac{y^k\dd y}{\sqrt{P_d(y)}}$ is holomorphic if $k = 0\ldots g-1$. However, if~ $k = g\ldots 2g-1$ a pole at infinity occurs and the differential is therefore~meromorphic.

\textls[-35]{A homology basis of closed integration paths $\{ a_i, b_i|i=1,...,g \}$ can be introduced to calculate  the period matrices  $(2\omega, 2\omega ')$ and $(2\eta, 2\eta ')$ of the solution $y(x)$ of the inversion problem~\eqref{eqn:hyper-firstkind}}
\begin{align}
	2\omega_{ij} &= \oint_{a_j} \dd u_i, \phantom{-}\qquad 2\omega'_{ij} = \oint_{b_j} \dd u_i, \nonumber\\ 
	2\eta_{ij} &= -\oint_{a_j} \dd r_i, \qquad 2\eta'_{ij} = -\oint_{b_j} \dd r_i .
	\label{eqn:periods}
\end{align}

The period matrices satisfy the Legendre relation, see~\cite{Hackmann:2008zz} or~\cite{Enolski:2011id}.
If the periods are calculated numerically for a specific problem, the~Legendre relation can be used to check the calculations.

Equation \eqref{eqn:hyper-firstkind} is closely related to the \emph{Jacobi inversion problem}, given by the equations
\begin{equation}
	\vec{x}=\sum\limits_{i=1}^{g} \int_{y_0}^{y_i} \dd \vec{u} \, .
\end{equation}

In the case $y_0=\infty$ it is possible to find a solution $\vec{y}$ for a given $\vec{x}$. Each component of the solution vector $\vec{y}$ is determined by the equation~\cite{Hackmann:2008zz}
\begin{equation}
	\frac{\lambda_{2g+1}}{4}y^g-\sum_{i=1}^g \wp_{gi}(\vec{x})y^{i-1} =0 \,
	\label{eqn:jacobiprob-sol}
\end{equation}
and can be extracted using the theorem of Vieta. However, it should be noted that the order of the components $y_i$ is not defined. The~generalized Weierstraß function is defined as the second logarithmic derivative of the Kleinian $\sigma$-function
\begin{equation}
\wp_{ij}(\vec{u}) = -\frac{\partial}{\partial u_i}\frac{\partial}{\partial u_j} \ln \sigma(\vec{u}) \, .
\end{equation}

The Kleinian $\sigma$-function is defined as~\cite{Hackmann:2008zz}
\begin{equation}
    \sigma (\vec{u} ) = C \mathrm{e}^{-\half \vec{u}^T \eta\omega^{-1} \vec{u}} \theta \left( (2\omega)^{-1}\vec{u} + \vec{K}_{x_0}; \tau\right)
\end{equation}
with $\tau = \omega^{-1}\omega'$. The~constant $C$ is given in~\cite{Enolski:2011id}. The~$\theta$-function is
\begin{equation}
    \theta (\vec{u}, \tau) = \sum_{\vec{m}\in\mathbb{Z}^g} \mathrm{e}^{i\pi\vec{m}^T(\tau m + 2\vec{u})} \, .
\end{equation} 
$\vec{K}_{x_0}$ is the vector of Riemann constants~\cite{Hackmann:2008zz, Enolski:2011id}.

The solution of Equation \eqref{eqn:hyper-firstkind} can be obtained in a limiting process which restricts the Jacobi inversion problem to the $\Theta$-divisor, the~set of zeros of the $\theta$-function. 
Let us demonstrate the limiting process in the case of genus 2 (see~\cite{Hackmann:2008zz, Hackmann:2008tu}). For~$g=2$ the Jacobi inversion problem is
\begin{align}
    x_1 &= \int_{y_0}^{y_1} \frac{\dd y}{\sqrt{P_5(y)}} +\int_{y_0}^{y_2} \frac{\dd y}{\sqrt{P_5(y)}}\nonumber\\
    x_2 &= \int_{y_0}^{y_1} \frac{y\dd y}{\sqrt{P_5(y)}} +\int_{y_0}^{y_2} \frac{y\dd y}{\sqrt{P_5(y)}} \, .
    \label{eqn:jacobiprob-gen2}
\end{align}

If $y_0=\infty$, the~Jacobi inversion problem has a solution in the form of Equation \eqref{eqn:jacobiprob-sol}. Therefore we rewrite the Equation \eqref{eqn:jacobiprob-gen2}
\begin{align}
    z_1 &= \int_\infty^{y_1} \frac{\dd y}{\sqrt{P_5(y)}} +\int_\infty^{y_2} \frac{\dd y}{\sqrt{P_5(y)}}\nonumber\\
    z_2 &= \int_\infty^{y_1} \frac{y\dd y}{\sqrt{P_5(y)}} +\int_\infty^{y_2} \frac{y\dd y}{\sqrt{P_5(y)}}
    \label{eqn:jacobiprob-gen2-infty}
\end{align}
with
\begin{equation}
  \vec{z} = \vec{x} - 2\int_{y_0}^\infty \dd \vec{u} \, .
\end{equation}

The solution of Equation \eqref{eqn:jacobiprob-gen2-infty} can by expressed as
\begin{align}
    y_1+y_2 &=\wp_{22}(\vec{z}) \nonumber\\
    y_1y_2 &= -\wp_{21}(\vec{z})
    \label{eqn:jacobipron-sol-gen2}
\end{align}
by applying the theorem of Vieta to Equation \eqref{eqn:jacobiprob-sol}.

Now we take the limit $y_2 \to \infty$. In~this limit one can write $y_1$ as
\begin{equation}
    y_1 = \lim_{y_2 \to \infty} \frac{y_1y_1}{y_1+y_2} \, .
\end{equation}

Inserting Equation \eqref{eqn:jacobipron-sol-gen2} yields
\begin{align}
     y_1 &= \lim_{y_2 \to \infty} -\frac{\wp_{21}(\vec{z})}{\wp_{22}(\vec{z})} \\
         &= \lim_{y_2 \to \infty} \frac{\sigma(\vec{z})\sigma_{12}(\vec{z})-\sigma_1(\vec{z})\sigma_2(\vec{z})}{\sigma_2^2(\vec{z}) - \sigma(\vec{z})\sigma_{22}(\vec{z})} \\
         &=\frac{\sigma(\vec{x}_\infty)\sigma_{12}(\vec{x}_\infty)-\sigma_1(\vec{x}_\infty)\sigma_2(\vec{x}_\infty)}{\sigma_2^2(\vec{x}_\infty) - \sigma(\vec{x}_\infty)\sigma_{22}(\vec{x}_\infty)}
\end{align}
where $\sigma_i$ is the $i$th derivative of the Kleinian $\sigma$-function and $\vec{x}_\infty=\lim_{y_2 \to \infty} \vec{z}$. It can be shown~\cite{Hackmann:2008zz} that $\vec{x}_\infty$ is an element of the $\Theta$-divisor, which is the set of zeros of the $\theta$-function. That means $\theta(\vec{x}_\infty)=0$ and therefore $\sigma(\vec{x}_\infty)=0$. Then we have
\begin{equation}
    y_1 = - \frac{\sigma_1(\vec{x}_\infty)}{\sigma_2(\vec{x}_\infty)} \, .
\end{equation}
In the end we want to find the inversion of the integral \eqref{eqn:hyper-firstkind}. For~this we have to consider
\begin{equation}
    \vec{x}_\infty=\lim_{y_2 \to \infty} \vec{z} = \lim_{y_2 \to \infty} \vec{x} - 2\int_{y_0}^\infty \dd \vec{u} = \int_{y_0}^{y_1} \dd \vec{u} - \int_{y_0}^\infty \dd \vec{u} \, .
\end{equation}

In a nutshell, for~a genus 2 curve the inversion of the integral \eqref{eqn:hyper-firstkind} with the initial value $y_{\mathrm in}=y(x_{\mathrm in})$ is
\begin{equation}
	y(x) = -\left. \frac{\sigma_1 (\vec{x}_\infty)}{\sigma_2 (\vec{x}_\infty)} \right| _{ \sigma (\vec{x}_\infty)=0} \, .
	\label{eqn:hyper-firstkind-solution}
\end{equation}

The vector $\vec{x}_\infty$ depends on the considered holomorphic integral:
\begin{equation}
	\vec{x}_\infty = 
	\left( 
	\begin{array}{c}
		x-x_\mathrm{in}'\\
		x_2 
	\end{array}
	\right) 
	\quad \text{to solve} \quad 
	x-x_\mathrm{in} = \int_{y_\mathrm{in}}^{y}\! \frac{\dd y'}{\sqrt{ P_5(y')}}
	\label{eqn:xinf-1}
\end{equation} 
and
\begin{equation}
	\vec{x}_\infty = 
	\left( 
	\begin{array}{c}
		x_1 \\
		x-x_\mathrm{in}''
	\end{array}
	\right) 
	\quad \text{to solve} \quad 
	x-x_\mathrm{in} = \int_{y_\mathrm{in}}^{y}\! \frac{y \, \dd y'}{\sqrt{ P_5(y')}} \, ,
	\label{eqn:xinf-2}
\end{equation}
where $x_\mathrm{in}'= x_\mathrm{in}+\int_{y_\mathrm{in}}^{\infty}\! \frac{\dd y'}{\sqrt{ P_5(y')}}$ and $x_\mathrm{in}''= x_\mathrm{in}+\int_{y_\mathrm{in}}^{\infty}\! \frac{y \, \dd y'}{\sqrt{ P_5(y')}}$. The~components $x_1$ and $x_2$ are determined by the condition $\sigma (\vec{x}_\infty)=0$.

In higher genera the inversion is given by~\cite{Enolski:2010if,Enolski:2011id}
\begin{equation}
	y(x)=-\left. \frac{
		\dfrac{\partial^{M+1}}{\partial x_1 \partial x_{g}^M }  \sigma(\vec{x}_\infty)}
	{\dfrac{\partial^{M+1}}{\partial x_2\partial x_{g}^M } \sigma(\vec{x}_\infty)} \right|_{\vec{x}_\infty \in \Theta_1}, \qquad M=\frac{(g-2)(g-3)}{2}+1 \, ,
	\label{eqn:inversion-higher-g}
\end{equation}
where
\begin{equation}
	{\Theta}_1 = \left\{\vec{x}_\infty \in \mathrm{Jac}(X_g)\; \Big|\; \sigma(\vec{x}_\infty)=0,\; \frac{\partial^j }{\partial x_g^j} \sigma(\vec{x}_\infty)=0 \quad \forall\,j=1,\ldots, g-2\right\} \, . 
\end{equation}

\sloppy{$\mathrm{Jac}(X_g) = \mathbb{C}_g / \Gamma$ is the Jacobian of the Riemannian surface $X_g$, where $\Gamma={\omega v + \omega'v'|v,v'\in \mathbb{Z}_g}$ is the lattice spanned by the periods $\omega_{ij}$ and $\omega'_{ij}$.
The solution formula \eqref{eqn:inversion-higher-g} is a conjecture based on the properties of the Schur-Weierstraß functions.
A relation similar to Equation \eqref{eqn:inversion-higher-g} holds for the Schur-Weierstraß polynomials and in~\cite{Enolski:2010if, Enolski:2011id} it was conjectured that this formula can also be used for the $\sigma$-function. An~analogue of this formula was considered in~\cite{Matsutani:2008}.
In the case genus $g=3$ Equation \eqref{eqn:inversion-higher-g} also holds, as~shown~in~\cite{Onishi:1998}.}

However, some geodesic equations yield \emph{hyperelliptic integrals of the third kind}
\begin{equation}
	\int_{y_\mathrm{in}}^{y} \frac{1}{y'-p} \frac{\dd y'}{\sqrt{ P_d(y')}}\, ,
	\label{eqn:hyper-thirdkind}
\end{equation}
\textls[-25]{where $p$ is a pole and $P_d(y)$ a polynomial of order $d$. A~formula to solve these integrals, which can be proven with the help of the Riemann vanishing theorem~\cite{Mumford:1983}, was found in~\cite{Enolski:2011id}}
\begin{align}
	\int_{y_\mathrm{in}}^y \frac{1}{y'-p}\frac{\dd y'}{\sqrt{ P_d(y')}} =  \frac{1}{\sqrt{ P_d(p)}} \left[ -2 \int_{y_\mathrm{in}}^y \dd\vec{u}^T  \int_{e_2}^p \dd\vec{r} \right.
	&+\ln \frac{\sigma\left(\int_{\infty}^y \dd\vec{u}- \int_{e_2}^{p} \dd \vec{u} - \vec{K}_\infty  \right)}{\sigma\left(\int_{\infty}^y \dd \vec{u}+ \int_{e_2}^{p} \dd \vec{u} - \vec{K}_\infty \right)}
	\nonumber\\
	& \left. 
	- \ln \frac{\sigma\left(\int_{\infty}^{y_\mathrm{in}} \dd \vec{u} - \int_{e_2}^{p} \dd \vec{u} - \vec{K}_\infty  \right)}{\sigma\left(\int_{\infty}^{y_\mathrm{in}} \dd\vec{u} + \int_{e_2}^{p} \dd \vec{u} - \vec{K}_\infty  \right)} \right]. 
	\label{eqn:sol-thirdkind}
\end{align}
$\dd\vec{u}$ and $\dd\vec{r}$ are the vectors of the holomorphic differentials of the first kind and the meromorphic differentials of the second kind respectively (Equation \eqref{eqn:hyper-differentials}). The~basepoint $e_2$ is a zero of the polynomial $P_d$. $\vec{K}_\infty$ is the vector of Riemann constants.
The integral $\int_{e_2}^p \dd\vec{r}$ can be rewritten in terms of the $\zeta$-function
\begin{align}
    \vec{\zeta}(\vec{u}) &= (\zeta_1(\vec{u}), \ldots, \zeta_g(\vec{u}))^T \\
    \zeta_i(\vec{u}) &= \frac{\partial}{\partial u_i} \ln \sigma (\vec{u})
\end{align}
and the characteristic
\begin{equation}
	[\mathfrak{A}_i] = \left(
	\begin{array}{c}
		\vec{\varepsilon}_i'^T\\
		\vec{\varepsilon}_i^T
	\end{array}
	\right) =
	\left(
	\begin{array}{c}
		\varepsilon_{i1}' \quad \varepsilon_{i2}'\\
		\varepsilon_{i1}  \quad \varepsilon_{i2}
	\end{array}
	\right) 
\end{equation}
of a branch point $e_i$ \cite{Enolski:2011id}. In~\cite{Enolski:2011id} the characteristics are explicitly calculated for $g=2$ in section (V.A.) and for $g=3$ in section (VI.A.). The~vectors $\vec{\varepsilon}_i$ and $\vec{\varepsilon}_i'$ $\in\mathbb{R}^2$ have the entries $\half$ or $0$. Then the integral is
\begin{equation}
	\int_{e_2}^p \dd\vec{r}= \vec{\zeta} \left( \int_{e_2}^p \dd \vec{u} + \vec{K}_\infty  \right) - 2( {\eta}^{\prime} \vec{\varepsilon}^\prime + {\eta}\vec{\varepsilon} )  - \half \vec{\mathfrak{Z}}(p,\sqrt{ P_d(p)})
\end{equation}
where $\eta$ and $\eta'$ are the half period matrices (see Equation \eqref{eqn:periods}), and~the $g$th component of the vector $\vec{\mathfrak{Z}}(Z,W)$ is $\mathfrak{Z}_g(Z,W)=0$ and for $1\leq j<g$ we have
\begin{equation}
	\mathfrak{Z}_j(Z,W)=\frac{W}{\prod_{k=2}^{g} (Z-e_{2k})}\sum_{k=0}^{g-j-1}(-1)^{g-k+j+1}Z^kS_{g-k-j-1}(\vec{e}) \, .
\end{equation}

The $S_k(\vec{e})$ are elementary symmetric functions of order $k$ built on  $g-1$ branch points $e_4,\ldots, e_{2g}$: $S_0=1$, $S_1=e_4+\ldots+e_{2g}$, etc~\cite{Enolski:2011id}.

\section{\boldmath $g=2$ \unboldmath Examples for Geodesic~Motion}

\subsection{Higher Dimensional Schwarzschild Black~Holes}

Schwarzschild black hole spacetimes in $D$ dimensions are given by~\cite{Tangherlini:1963bw}
\begin{equation}
    ds^2 = - f(r) dt^2 + f(r)^{-1} dr^2 + r^2 d\Omega^2_{D-2} \ , \ \ \ f(r) = 1-\left(\frac{r_S}{r}\right)^{D-3} \ ,
\end{equation}
with Schwarzschild radius $r_S$ and standard metric on the $D-2$-sphere $d\Omega^2_{D-2}$.
\textls[-25]{Since energy $E$ and angular momentum $L$ are conserved, and~the motion is confined to an equatorial plane}
\begin{equation}
E= f(r) \frac{dt}{d\tau} \ , \qquad L = r^2\frac{d\varphi}{d\tau} \ ,
\end{equation}
the remaining equation of motion is the radial equation
\begin{equation}
    \left(\frac{dr}{d\tau}\right)^2 = E^2 - f(r) \left( \frac{L^2}{r^2} -\delta \right) \ , \label{drds}
\end{equation}
respectively, the~orbital equation
\begin{equation}
    \left(\frac{dr}{d\phi}\right)^2 = \frac{r^4}{L^2} \left(E^2 - \left(1 - f(r)\right) \left( \frac{L^2}{r^2} - \delta \right) \right) \ , \label{EOMrphi}
\end{equation}
yielding the particle orbit in the black hole spacetime.
Introducing dimensionless quantities
\begin{equation}
\lambda=\frac{r^2_{S}}{L^2} \, , \quad \mu = E^2 \, , \qquad \tilde{r} = \frac{r}{r_{\rm S}} \,
\end{equation}
the right hand side of Equation~(\ref{EOMrphi}) can be expressed as $P_n(\tilde r)/\tilde r^m$, where $P_n(\tilde r)$ is a polynomial of order $n$.
Moreover, a~substitution is possible in odd dimensions, $u=1/\tilde r^2$, to~reduce the order of the polynomial by a factor of 2. 
Thus a $g=2$ case is obtained in $D=6$, 9 and 11 dimensions (while $D=5$ and 7 are still elliptic) \cite{Hackmann:2008tu}.
In 9 and 11 dimensions, the~orbital equations read
\begin{eqnarray}
 \left(\frac{d u}{d\varphi}\right)^2 &=&4 u \left(u^4 + \lambda u^3 - u + \lambda(\mu-1)\right) = 4 P_5(u) \ , \\
 \left(\frac{d u}{d\varphi}\right)^2 &=& 4 u \left(u^5 + \lambda u^4 - u + \lambda(\mu-1)\right) = 4 P_6(u) \ ,
\end{eqnarray}
respectively, leading to the solutions
\begin{eqnarray}
 \tilde{r}(\varphi) &= & \sqrt{- \frac{\sigma_2(\vec \varphi_{\infty, z_1})}{\sigma_1(\vec \varphi_{\infty, z_1})} } \ , \\
 \tilde{r}(\varphi) &=&  \frac{1}{\sqrt{- \frac{\sigma_2(\vec \varphi_{\infty, z_2})}{\sigma_1(\vec \varphi_{\infty, z_2})} + u_6}} \ ,
\end{eqnarray}
where in the latter 11 dimensional case a substitution $u=\frac{1}{x} + u_6$ was performed, with~$u_6$ a root of $P_6(u)$, transforming the orbital equation to $\left(x\frac{dx}{d\varphi}\right)^2= 4 P_5(x)$ \cite{Hackmann:2008tu}.
The $\vec \varphi_{\infty, z_i}$ are defined as
\begin{eqnarray}
      \vec \varphi_{\infty, z_2} & =& \begin{pmatrix} \int_{x_{\rm in}}^{x_1} dz_1 - \int_{x_{\rm in}}^{\infty} dz_1 \\ \varphi_{z_2} - \varphi^\prime_{{\rm in}, z_2}  \end{pmatrix}  \ , \\
      \vec \varphi_{\infty, z_1} & =& \begin{pmatrix} \varphi_{z_1} - \varphi^\prime_{{\rm in}, z_1}  \\ \int_{x_{\rm in}}^{x_1} dz_2 - \int_{x_{\rm in}}^{\infty} dz_2 \end{pmatrix}  \ ,
\end{eqnarray}
with $\varphi^\prime_{{\rm in}, z_2} = \varphi_{\rm in} + \int^\infty_{x_{\rm in}} d z_2$ and $\varphi^\prime_{{\rm in}, z_1} = \varphi_{\rm in} + \int^\infty_{x_{\rm in}} d z_1$.

Further analysis of the possible particle motion reveals, that these higher dimensional Schwarzschild black hole spacetimes do not allow for periodic bound orbits. 
Only escape orbits away from the black hole to spatial infinity are allowed and orbits terminating at the central black hole~singularity.

\subsection{Rotating Dyonic $U(1)^2$ Black Holes}

The problem of dark matter and dark energy is still an unsolved problem of physics, which could possibly be solved by introducing scalar fields like the dilaton and the axion. An~interesting spacetime containing these fields is the rotating dyonic black hole with four electromagnetic charges of the $U(1)^2$ gauged supergravity found by Chow and Compère~\cite{Chow:2013gba}. The~exact solutions of the equations of motion in the rotating dyonic black hole spacetime were found in~\cite{Flathmann:2016knq}. The~metric is given by
\begin{adjustwidth}{-\extralength}{0cm}
\begin{equation}
	\dd s^2 = -\frac{R_g}{B-aA}\left(\dd t -\frac{A}{\Xi}\dd \phi\right)^2 + \frac{B-aA}{R_g}\dd r^2  
	+ \frac{\Theta_g a^2 \sin^2{\vartheta}}{B-aA}\left(\dd t-\frac{B}{a\Xi}\dd \phi\right)^2 + \frac{B-aA}{\Theta_g} \dd \vartheta^2 \, ,
\end{equation}
\end{adjustwidth}
with
\begin{equation}
 \begin{split}
  R_g &= r^2-2Mr+a^2+e^2-N_g^2+g^2[r^4 
      +(a^2+6N_g^2-2v^2)r^2+3N_g^2(a^2-N_g^2)]\\
  \Theta_g &= 1-a^2g^2\cos^2{\vartheta}-4a^2N_g \cos{\vartheta}\\
  A &= a\sin^2{\vartheta}+4N_g\sin^2{\frac{\vartheta}{2}}\\
  B &= r^2+(N_g+a)^2-v^2 \\ 
 \Xi &= 1-4N_g ag^2-a^2g^2 \, .
 \end{split}
\end{equation}

Here $M$ describes the mass of the black hole, $a$ is the rotation parameter, $e$ and $v$ correspond to the charges, $N_g$ is the Newman-Unti-Tamburino (NUT) parameter and $g$ is the gauge coupling constant. The~ Boyer-Lindquist like coordinates $(t, r, \vartheta, \phi)$ transform to Cartesian coordinates as
\begin{equation}
 	\begin{split}
		x&=\sqrt{(r^2+a^2)}\sin{\vartheta}\cos{\varphi} \\
		y&=\sqrt{(r^2+a^2)}\sin{\vartheta}\sin{\varphi} \\
		z&=r\cos{\vartheta} \, .
	\end{split}
\end{equation}

Using the Hamilton-Jacobi formalism as described in Section~\ref{sec:GeoMotion} one finds four differential equations which describe the motion of particles and light in the above spacetime.
\begin{eqnarray}
 \left(\frac{\dd \tr}{\dd \gamma}\right)^2 &=& X \, , \label{eqn:dyon-r-equation} \\
 \sin^2{\vartheta}\left(\frac{\dd \vartheta}{\dd \gamma}\right)^2 &=& Y \, , \label{eqn:dyon-theta-equation} \\
\left(\frac{\dd \phi}{\dd \gamma}\right) &=& \frac{\ta\tXi(\tB E -\ta \tL \tXi)}{\tR} +\frac{\tXi (\tA E -\tL \tXi)}{\tTh \sin^2{\vartheta}} \, , \label{eqn:dyon-phi-equation} \\
\left(\frac{\dd \tlt}{\dd \gamma}\right) &=& \frac{\tB(\tB E -\ta\tL\tXi)}{\tR} +\frac{\tA(\tL\tXi-\tA E)}{\tTh \sin^2{\vartheta}} \, . \label{eqn:dyon-t-equation}
\end{eqnarray}
with the functions
\begin{equation}
\begin{split}
  X &= (\tB E -\ta\tL\tXi)^2+\tR(\tK -\tB\delta) \, , \\
  Y &= -(\tA E-\tL\tXi)^2+\tTh\sin^2{\vartheta}(\ta\tA\delta-\tK) \, ,\\
  \tR &= \tr^2-\tr +\ta^2+\te^2-\tN^2 \tg^2[\tr^4+(\ta^2+6\tN^2-2\tv^2)\tr^2+3\tN^2(\ta^2-\tN^2)] \, ,\\
  \tTh &= 1 -\ta^2\tg^2\cos^2{\vartheta}+4\ta^2\tg^2\tN\cos{\vartheta} \, , \\
  \tA &= \ta^2 \sin^2{\vartheta}+2 \tN (1-\cos{\vartheta}) \, , \\
  \tB &= \tr^2+(\tN+\ta)^2-\tv^2 \, ,\\
  \tXi &= 1-\ta^2\tg^2-4\ta\tN\tg^2 \, .
\end{split}
\end{equation}

Here dimensionless quantities were used
\begin{equation}
 \begin{split}
	\tr&=\frac{r}{2M} \ , 
	\,\, \tlt=\frac{t}{2M} \ , 
	\,\, \tilde{\tau}=\frac{\tau}{2M} \ , 
	\,\, \tN=\frac{N_g}{2M} \ ,
	\,\, \ta=\frac{a}{2M} \ , \\
	\,\, \tg&=2Mg \ , 
	\,\, \te=\frac{e}{2M}  \ ,
	\,\, \tv=\frac{b}{2M}  \ ,
	\,\, \tK=\frac{K}{2M}  \ ,
	\,\, \tL=\frac{L}{2M} \ .
\end{split}
\end{equation}

The definition of $\gamma$ with $\dd \tilde{\tau}=(\tB-\ta \tA)\dd\gamma$ simplifies the equation by absorbing the $r$ and $\vartheta$ dependent prefactor $(\tB-\ta \tA)$. 

The equations of motion are of hyperelliptic type and can be solved in terms of the Kleinian $\sigma$-function. The~$r$-Equation \eqref{eqn:dyon-r-equation} yields a hyperelliptic integral of the first kind. In~general the right hand side of \eqref{eqn:dyon-r-equation} is a polynomial of order six $X = \sum_{i=1}^6 a_i\tr^i$. The~substitution $\tr=\pm \frac{1}{x}+\tr_0$, where $\tr_0$ is a zero of $X$, transforms $X$ into a polynomial of order five and the $r$-equation becomes
\begin{equation}
 \left(x\frac{\dd x}{\dd \gamma}\right)^2=\sum_{i=0}^5 b_ix^i=:P_5^{\tr}(x) \, .
\end{equation}

A separation of variables yields the hyperelliptic integral
\begin{equation}
    \gamma - \gamma_{\mathrm in} = \int _{x_{\mathrm in}}^x \frac{x'\dd x'}{\sqrt{P_5^{\tr}(x')}} .
    \label{eqn:dyon-hyperint} 
\end{equation} 

As described in Section~\ref{sec:hyperell}, see Equations \eqref{eqn:hyper-firstkind-solution} and \eqref{eqn:xinf-2}, the~solution of the above equation is
\begin{equation}
 x= -\frac{\sigma_1(\vec{\gamma}_{\infty})}{\sigma_2(\vec{\gamma}_{\infty})} \, , 
\end{equation}
where $\sigma_i$ is the $i$th derivative of the Kleinian $\sigma$-function and
\begin{equation}
 \vec{\gamma}_{\infty} = \left(\gamma_1, \gamma - \gamma_{\rm in}- \int_{x_{\rm in}}^{\infty}{{\frac{x \dd x}{\sqrt{P_5^{\tr}(x)}}}}\right)^T \, .
 \label{eqn:dyon-gammainf}
\end{equation}
\textls[-25]{$\gamma_1$ is determined by the condition $\sigma (\vec{\gamma}_\infty)=0$.
A resubstitution yields the full solution of \eqref{eqn:dyon-r-equation}}
\begin{equation}
 \tr(\gamma)= \mp \frac{\sigma_2(\vec{\gamma}_{\infty})}{\sigma_1(\vec{\gamma}_{\infty})}+\tr_0 \, .
 \label{eqn:dyon-rsol}
\end{equation}

The  $\vartheta$-Equation \eqref{eqn:dyon-theta-equation} can be solved similarly by substituting $\cos{\vartheta}=\pm\frac{1}{\nu}+\nu_{0}$, where $\nu_{0}$ is a zero of $Y$, so that
\begin{equation}
  \left(\nu\frac{\dd \nu}{\dd \gamma}\right)^2=\sum_{i=0}^5 b'_i\nu^i=:P_5^{\vartheta}(\nu) \, .
\end{equation}

The solution is
\begin{equation}
\vartheta = \arccos{\left(\mp \frac{\sigma_2(\vec{\gamma'}_{\infty})}{\sigma_1(\vec{\gamma'}_{\infty})}+ \nu_0\right)} \, 
\end{equation}
with
\begin{equation}
 \vec{\gamma'}_{\infty} = \left(\gamma_1', \gamma - \gamma_{\rm in}- \int_{\nu_{in}}^{\infty}{{\frac{\nu \dd \nu}{\sqrt{P_5^{\vartheta}(\nu)}}}}\right)^T \, ,
\end{equation}
where $\gamma_1'$ is determined by the condition $\sigma (\vec{\gamma'}_\infty)=0$.

The $\phi$-Equation \eqref{eqn:dyon-phi-equation} involves hyperelliptic integrals of the third kind. Here we use the Equations \eqref{eqn:dyon-r-equation} and \eqref{eqn:dyon-theta-equation} to rewrite the $\phi$-Equation \eqref{eqn:dyon-phi-equation} as
\begin{equation}
 \phi -\phi_{in} = \int_{\tr_{\rm in}}^{\tr}{\frac{\ta\tXi(\tB E-\ta\tL\tXi)}{\tR}\frac{\dd \tr}{\sqrt{X}}} + \int_{\vartheta_{\rm in}}^{\vartheta}{\frac{\tXi(\tA E-\tL\tXi)}{\tTh\sin{\vartheta}}\frac{\dd \vartheta}{\sqrt{Y}}} = I_{\tr}(\tr)+I_{\vartheta}(\vartheta) \, .
\end{equation}
$I_{\tr}$ and $I_{\vartheta}$ can be solved separately. Both integrals can be decomposed into several integrals of the form
\begin{equation}
 I= \int_{x_{\rm in}}^{x} \frac{\dd x'}{(x-Z)\sqrt{P_5(x)}} \, ,
\end{equation}
where $Z$ is a pole and $P_5(x)$ is a polynomial of the fifth order. The~solution is (see Equation \eqref{eqn:sol-thirdkind} in Section~\ref{sec:hyperell})
\begin{adjustwidth}{-\extralength}{0cm}
\begin{equation}
 I=\frac{2}{\sqrt{P_5{(Z)}}}\int_{x_{\rm in}}^x{\dd\vec{z}^T}\int_{e_2}^{Z}{\dd\vec{y}}
+\ln{\left(\frac{\sigma(\int_{\infty}^x{\dd\vec{z}}-\int_{e_2}^Z{\dd\vec{z}})}
{\sigma(\int_{\infty}^x{\dd\vec{z}}+\int_{e_2}^Z{\dd\vec{z}})}\right)}
-\ln{\left(\frac{\sigma(\int_{\infty}^{x_{\rm in}}{\dd\vec{z}}-\int_{e_2}^Z{\dd\vec{z}})}
{\sigma(\int_{\infty}^{x_{\rm in}}{\dd\vec{z}}+\int_{e_2}^Z{\dd\vec{z}})}\right)} \, .
\end{equation}
\end{adjustwidth}
$e_2$ is a zero of the polynomial $P_5(x)$ with the coefficients $a_k$ and the holomorphic and meromorphic differentials are
\begin{equation}
 \dd\vec{z}:=\left(\frac{\dd x}{\sqrt{P_5(x)}},\frac{x\dd x}{\sqrt{P_5(x)}}\right)^T
\end{equation}
\begin{equation}
 \dd\vec{y}=\left(\sum_{k=1}^4k a_{k+1}\frac{x^k\dd x}{4\sqrt{P_5(x)}},
\sum_{k=2}^3(k-1)a_{k+3}\frac{x^k\dd x}{4\sqrt{P_5(x)}}\right)^T \, .
\end{equation}

Analogously, the~solution of the $\tlt$-Equation \eqref{eqn:dyon-t-equation} can be~found.\\

\begin{figure}[H]

\begin{tabular}{ccc} 
        {\includegraphics[width=0.32\linewidth]{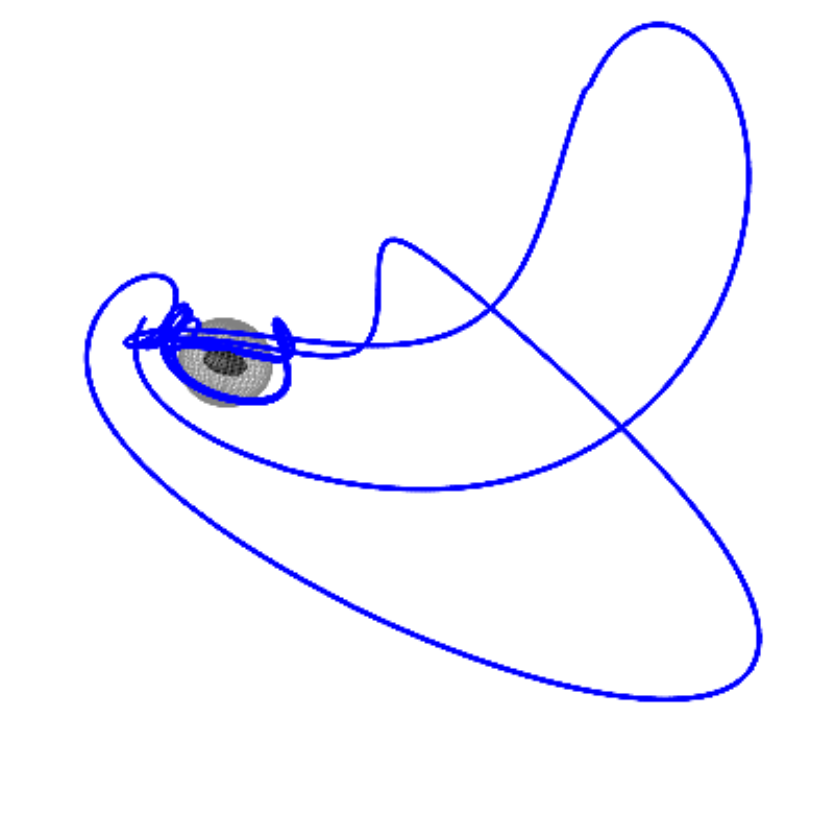}} & 
        
         {\includegraphics[width=0.32\linewidth]{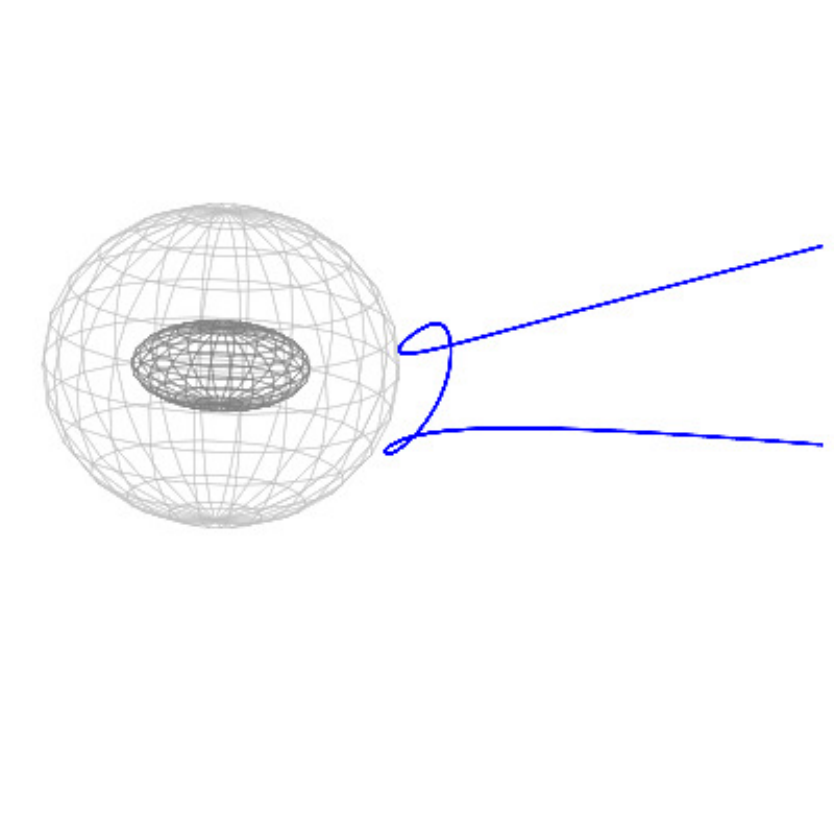}}   & 
         
          {\includegraphics[width=0.32\linewidth]{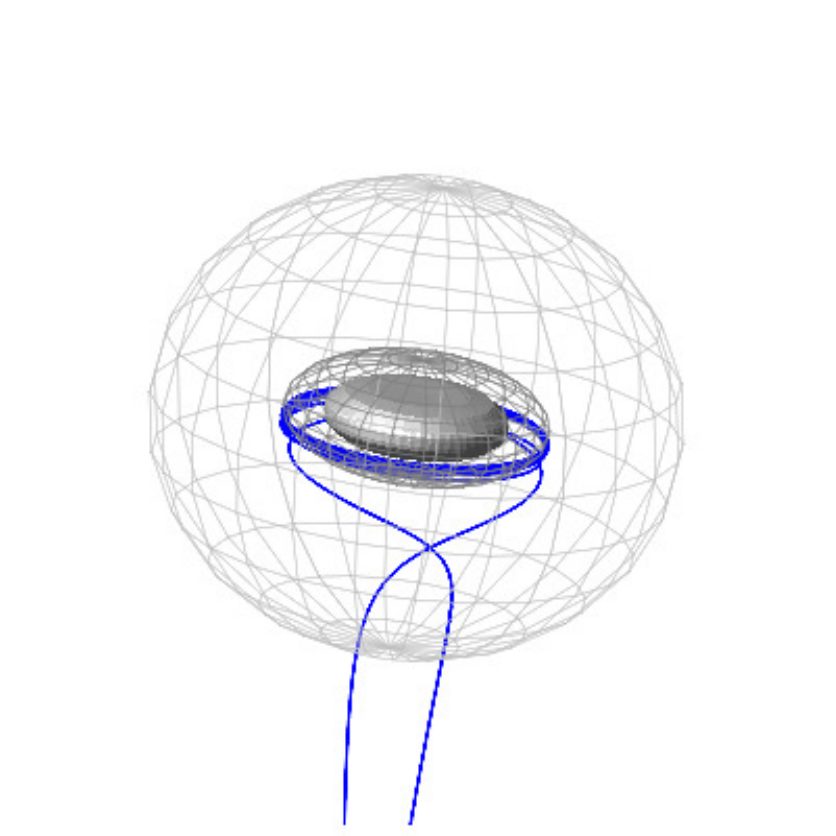}}  \\
   ({\bf a})&({\bf b})& ({\bf c})\\
\end{tabular}

	\caption{Orbits of particles (blue curves) around the rotating dyonic $U(1)^2$ black hole. The~horizons are represented by grey ellipsoids and the dark grey structure is the singularity of the black~hole. (\textbf{a}) Bound orbit; (\textbf{b}) Escape orbit; (\textbf{c}) Two-world escape orbit crossing both horizons twice and emerging into another universe.}
	\label{fig:dyon-orbits}
\end{figure}

With the full set of analytical solutions we can visualize the orbits in this spacetime, see Figure~\ref{fig:dyon-orbits}. 

For the calculation of a specific orbit, first the periods have to be computed according to Equation \eqref{eqn:periods}. The~integration path depends on the location of the zeros of the polynomial in the complex plane. The~zeros are singularities of the integrand and therefore the integral has to be split into several parts, integrating from one zero to the next. When complex zeros occur, real and imaginary parts are integrated separately. One has to keep in mind that the sign of the square root $\sqrt{P_d}$ is different for each part of the integration, see~\cite{Hackmann:2008zz} for further details and pictures of the integration~paths.

As the starting point $x_{\rm in}$ in Equation \eqref{eqn:dyon-hyperint} we choose one of the turning points of the orbit we want to plot, i.e.,~a zero of the polynomial. This has the advantage, that $\gamma - \gamma_{\rm in}- \int_{x_{\rm in}}^{\infty}{{\frac{x \dd x}{\sqrt{P_5^{\tr}(x)}}}}$ in Equation \eqref{eqn:dyon-gammainf} can be expressed in terms of the periods (if we choose $\gamma_{\rm in}=0$) .
The solution of the radial equation is computed pointwise. For~each gamma in {$\vec{\gamma}_\infty$}, we have to compute a $\gamma_1$ according to the condition {$\sigma (\vec{\gamma}_\infty)=0$}. Here we use the Newton-Raphson method to determine $\gamma_1$. With~the resulting vector {$\vec{\gamma}_\infty$}, we can compute the solution $r(\gamma)$ (Equation \eqref{eqn:dyon-rsol}). All our calculations are performed with the help of the software MAPLE {, including the numerical calculation of the $\theta$- and $\sigma$-function and their derivatives}.

Possible types of motion are bound orbits and escape orbits, which can also cross the horizons. The~maximal analytic extension of the spacetime gives rise to an infinite set of \textit{universes} or \textit{worlds},
i.e.,~asymptotically flat regions, that are connected by intermediate regions delimited by horizons and by regions containing a singularity. Thus a particle can pass from one asymptotically flat region via an outer horizon into an intermediate region, from~there via an inner horizon into a region with a singularity, leave again this region via an inner horizon, pass through another intermediate region, and~then leave across an outer horizon into another asymptotically flat region. Such a two-world escape orbit is illustrated in Figure~\ref{fig:dyon-orbits}c. Here the first outer horizon represents a black hole horizon for the particle, whereas the second outer horizon is experienced by the particle as a white hole horizon, allowing it to leave the black hole it had entered. The~Figure~\ref{fig:dyon-orbits}c does not distinguish between the universes, though, and~identifies their spatial coordinates.  Furthermore, the~shape of the singularity (which varies from ring-like to three dimensional structures) allows some geodesics to pass the singularity and reach negative $\tr$, that can be interpreted as reaching a \textit{universe} with anti-gravity. The~types of orbits around the rotating dyonic $U(1)^2$ black hole are similar those in the Kerr-Newman-AdS~spacetime.

\unskip

\subsection{Black~Rings}

In four dimensions the event horizon of a black hole is spherical, however, in~higher dimension new topologies arise. In~2001 Emparan and Reall~\cite{Emparan:2001wn}  presented a black hole solution with a topology of $S^1 \times S^2$: the rotating black ring. A~doubly spinning black
ring with two independent angular momenta was found by Pomeransky and Sen’kov~\cite{Pomeransky:2006bd}. Elvang constructed a charged singly spinning black ring in 2003~\cite{Elvang:2003yy} and soon after, Hoskisson presented a
doubly spinning version of the charged black ring~\cite{Hoskisson:2008qq}. Exact solutions of the equations of motion were found in the singly spinning black ring spacetime~\cite{Grunau:2012ai} and the charged doubly spinning black ring spacetime~\cite{Grunau:2012ri}.

The metric of a doubly spinning charged black ring can be written as
\begin{adjustwidth}{-\extralength}{0cm}
\begin{align}
 \mathrm{d} s^2 &= - D(x,y)^{-2/3}\frac{H(y,x)}{H(x,y)} (\mathrm{d}t + c\Omega)^2 + D(x,y)^{1/3}\frac{R^2H(x,y)}{(x-y)^2(1-\nu)^2} \left[ \frac{\mathrm{d}x^2}{G(x)} - \frac{\mathrm{d}y^2}{G(y)} \right. \nonumber \\ 
 & \left. + \frac{A(y,x)\mathrm{d}\phi^2 - 2L(x,y)\mathrm{d}\phi\mathrm{d}\psi - A(x,y)\mathrm{d}\psi^2}{H(x,y)H(y,x)} \right] \, .
\end{align}
\end{adjustwidth}

 The metric functions read
 \begin{adjustwidth}{-\extralength}{0cm}
\begin{align}
 G(x) &= (1-x^2)(1+\lambda x+\nu x^2) \nonumber \\
 H(x,y) &= 1+ \lambda^2 -\nu^2 + 2\lambda\nu (1-x^2)y + 2x\lambda (1-y^2\nu^2) + x^2y^2\nu (1-\lambda^2-\nu^2) \nonumber \\
 L(x,y) &= \lambda \sqrt{\nu} (x-y)(1-x^2)(1-y^2)[1+\lambda^2-\nu^2+2(x+y)\lambda\nu -xy\nu (1-\lambda^2-\nu^2)] \nonumber \\
 A(x,y) &= G(x)(1-y^2) [((1-\nu^2)-\lambda^2)(1+\nu)+y\lambda (1-\lambda^2 + 2\nu -3\nu^2)] \nonumber \\
        & + G(y)[2\lambda^2 + x\lambda((1-\nu)^2+\lambda^2) + x^2((1-\nu)^2-\lambda^2)(1+\nu) \nonumber \\
        & + x^3\lambda (1-\lambda^2-3\nu^2+2\nu^3) + x^4\nu (1-\nu)(1-\lambda^2-\nu^2)] \nonumber \\
 D(x,y)&=c^2-s^2\frac{H(y,x)}{H(x,y)} = 1+s^2\frac{2\lambda(1-\nu)(x-y)(1-\nu xy)}{H(x,y)} \, .
\end{align}
\end{adjustwidth}

The shape, mass and angular momenta of the black ring are represented by the parameters $R$, $\lambda$ and $\nu$, where $0\leq \nu <1$ and $2\sqrt{\nu} \leq \lambda <1+\nu$. The~metric reduces to a singly spinning black ring for $\nu=0$. The~doubly spinning black ring possesses two independent angular momenta and thus the rotation is given by
\begin{align}
 \Omega &=\Omega _\psi \mathrm{d}\psi + \Omega _\phi \mathrm{d}\phi \, ,
\end{align}
with
\begin{adjustwidth}{-\extralength}{0cm}
\begin{align}
 \Omega _\psi &= - \frac{R\lambda\sqrt{2((1+\nu)^2-\lambda^2)}}{H(y,x)}\frac{1+y}{1-\lambda +\nu} (1+\lambda -\nu + x^2y\nu (1-\lambda-\nu) + 2\nu x(1-y)) \, , \nonumber\\
 \Omega _\phi &= - \frac{R\lambda\sqrt{2((1+\nu)^2-\lambda^2)}}{H(y,x)}(1-x^2)y\sqrt{\nu} \, .
\end{align}
\end{adjustwidth}

The charge is described by the parameters
\begin{equation}
 c=\cosh (\alpha) \quad \text{and} \quad s=\sinh (\alpha) \quad \text{with} \quad \alpha \in \mathbb{R} \, .
\end{equation}

For  $c=1$ and $s=0$ one obtaines an uncharged black~ring.\\

The black ring metric is given in toroidal coordinates with $-1 \leq x \leq 1$, $-\infty < y \leq -1$ and $-\infty < t < \infty$.  $\phi$ and $\psi$ are $2\pi$-periodic. The~toroidal coordinates can be seen as two pairs of polar coordinates
\begin{equation}
\begin{array}{l} 
x_1=r_1 \sin(\phi)\\
x_2=r_1 \cos(\phi)
\end{array}
\quad\text{and}\quad
\begin{array}{l} 
x_3=r_2 \sin(\psi)\\
x_4=r_2 \cos(\psi)
\end{array}
\end{equation}
with
\begin{equation}
 r_1=R\frac{\sqrt{1-x^2}}{x-y} \quad\text{and}\quad r_2=R\frac{\sqrt{y^2-1}}{x-y}
\end{equation}
$x_1$, $x_2$, $x_3$, $x_4$ are four-dimensional Cartesian-like coordinates.

\textls[-25]{A ring-like curvature singularity is located at $y=-\infty$, light and particles cannot return from this area. At~$G(y)=0$ the metric has a coordinate singularity resulting in two~horizons}
\begin{eqnarray}
 y_{h+} &=& \frac{-\lambda + \sqrt{\lambda ^2 -4\nu}}{2\nu} \\
 y_{h-} &=& \frac{-\lambda - \sqrt{\lambda ^2 -4\nu}}{2\nu} \, .
\end{eqnarray}

If the angle $\phi$ is constant, the~black ring horizon has a donut-like topology $S^1\times S^1$. On~the other hand if $\psi$ is constant the horizon will look like two $S^2$ spheres.\\

To obtain the equations of motion for light and test particles in the five-dimensional spacetime of a charged doubly spinning black ring, we use the Hamilton-Jacobi formalism, see Section~\ref{sec:GeoMotion}. The~metric of the black ring and the Hamiltonian  $\mathscr{H} = \frac{1}{2}g^{ab}p_ap_b$ do not depend on the coordinates $t$, $\phi$ and $\psi$ and therefore three conserved momenta $p_a=g_{ab}\dot{x}^b$ with the associated killing vector fields $\partial / \partial t$, $\partial / \partial \phi$ and $\partial / \partial \psi$ exist
\begin{align}
 p_t &= -D(x,y)^{-2/3}\frac{H(y,x)}{H(x,y)} (\dot{t}+c\Omega _\phi \dot{\phi} +c\Omega _\psi \dot{\psi}) \equiv -E \\
 p_\phi &= -c\Omega _\phi E -D(x,y)^{1/3}\frac{R^2}{H(y,x)(x-y)^2(1-\nu)^2}(-A(y,x)\dot{\phi}+L(x,y)\dot{\psi}) \equiv \Phi \\
 p_\psi &= -c\Omega _\psi E -D(x,y)^{1/3}\frac{R^2}{H(y,x)(x-y)^2(1-\nu)^2}(A(x,y)\dot{\psi}+L(x,y)\dot{\phi}) \equiv \Psi \, .
\end{align}

The dot denotes a derivative with respect to the affine parameter $\tau$.

\textls[-35]{In the Hamilton-Jacobi equation we need the non-vanishing components of the inverse~metric}
\begin{equation}
 \begin{split}
   g^{tt} &= -D(x,y)^{2/3}\frac{H(x,y)}{H(y,x)}+c^2D(x,y)^{-1/3}\frac{(x-y)^2}{R^2H(x,y)} \\
   &\times\left[ \frac{\Omega_\phi^2A(y,x) - 2\Omega_\phi\Omega_\psi L(x,y) - \Omega_\psi^2 A(x,y)}{G(x)G(y)} \right]  \\
   g^{t\phi} &= cD(x,y)^{-1/3}\frac{(x-y)^2}{R^2H(x,y)} \frac{\Omega_\psi L(x,y) - \Omega_\psi A(x,y)}{G(x)G(y)} \\
   g^{t\psi} &= cD(x,y)^{-1/3}\frac{(x-y)^2}{R^2H(x,y)} \frac{\Omega_\phi L(x,y) + \Omega_\psi A(y,x)}{G(x)G(y)} \\
   g^{\phi\phi} &= D(x,y)^{-1/3}\frac{(x-y)^2}{R^2H(x,y)} \frac{A(x,y)}{G(x)G(y)}\\
   g^{\psi\psi} &= -D(x,y)^{-1/3}\frac{(x-y)^2}{R^2H(x,y)} \frac{A(y,x)}{G(x)G(y)}\\
   g^{\phi\psi} &= -D(x,y)^{-1/3}\frac{(x-y)^2}{R^2H(x,y)} \frac{L(x,y)}{G(x)G(y)}\\
   g^{xx} &= D(x,y)^{-1/3}\frac{(x-y)^2 (1-\nu)^2}{R^2H(x,y)} G(x) \\
   g^{yy} &= -D(x,y)^{-1/3}\frac{(x-y)^2 (1-\nu)^2}{R^2H(x,y)} G(y) \, .
 \end{split}
\end{equation}

$E$ is the energy of the particle and its angular momenta in $\phi$- and $\psi$-direction are $\Phi$ and $\Psi$. In~the $x$- and $y$-direction the conjugate momenta are:
\begin{align}
 p_x &= D(x,y)^{1/3}\frac{R^2H(x,y)\dot{x}}{(x-y)^2(1-\nu)^2G(x)} \\
 p_y &= -D(x,y)^{1/3}\frac{R^2H(x,y)\dot{y}}{(x-y)^2(1-\nu)^2G(y)} 
\end{align}

It is useful to split the polynomials $A(x,y)$ and $L(x,y)$ into $x$- and $y$-parts
\begin{align}
 A(x,y) &= G(x)\alpha (y) + G(y)\beta (x) \nonumber \\
 L(x,y) &= G(x)\delta(y) - G(y)\delta (x) \, ,
\end{align}
with
\begin{align}
 \alpha (\xi) &= \nu (1-\xi^2)[-(1+\lambda^2)-\nu (1-\nu)+\lambda\xi (2-3\nu) - (1-\lambda^2)\xi^2] \nonumber \\
 \beta (\xi) &= (1+\lambda^2) + \lambda\xi (1+(1-\nu)^2) - \nu\xi^2(2\lambda^2+\nu (1-\nu)) - \lambda\nu^2\xi^3(3-2\nu) \nonumber \\
             & -\nu^2\xi^4(1-\lambda^2+\nu(1-\nu)) \nonumber \\
 \delta (\xi) &= \lambda\sqrt{\nu}(1-\xi^2)(\lambda-(1-\nu^2)\xi -\lambda\nu\xi^2) \, .
\end{align}

Having the three constants of motion $E$, $\Phi$ and $\Psi$ one can think of an ansatz to solve the Hamilton-Jacobi equation
\begin{equation}
 S(\tau, t, x, y, \phi, \psi) = \frac{1}{2}\delta\tau -Et +\Phi\phi +\Psi\psi + S_x(x)+S_y(y) .
\end{equation}

Inserting everything into the Hamilton-Jacobi equation yields
\begin{adjustwidth}{-\extralength}{0cm}
\begin{equation}
 \begin{split}
 0 &= \delta -D(x,y)^{2/3}\frac{H(x,y)}{H(y,x)} E^2 + D(x,y)^{-1/3}\frac{(x-y)^2(1-\nu)^2}{R^2H(x,y)} \left[ G(x) \left( \frac{\partial S}{\partial x} \right)^2 
   - G(y) \left( \frac{\partial S }{\partial y} \right)^2  \right. \\
   & \left. + \frac{ A(x,y) \left(\Phi +c\Omega _\phi E\right)^2 -2L(x,y) \left(\Phi +c\Omega _\phi E\right) \left(\Psi +c\Omega _\psi E\right) 
   -A(y,x) \left(\Psi +c\Omega _\psi E\right)^2 }{(1-\nu)^2G(x)G(y)} \right] \, .
 \label{eqn:hj-c-d-ring}
 \end{split}
\end{equation}
\end{adjustwidth}

In general Equation \eqref{eqn:hj-c-d-ring} does not seem to be separable. However, it can be separated in three~cases:
\begin{enumerate}
    \item $E=\delta=0$: This special case describes zero energy null geodesics, which are realistic inside the ergoregion only.
    \item $x=\pm 1$: This case describes geodesics in the equatorial plane of the black ring, which is also the ``axis'' of rotation in $\phi$-direction. The~equatorial plane can be divided into two parts: The plane enclosed by the black ring $x=+1$ and the plane around the black ring $x=-1$
    \item $y=-1$: Here geodesics on the ``axis'' of rotation in $\psi$-direction are considered. The~case $y=-1$ describes a plane between two $S^2$ spheres which represent the horizon of the black ring.
\end{enumerate}

In the first case $E=\delta=0$ the Hamilton-Jacobi formalism yields five equations of motion, which are of elliptic type and can be solved in terms of the Weierstraß $\wp$-, $\sigma$- and $\zeta$-function. In~the second and third case, the~motion takes place in a plane and we get three equations of motion, which are of hyperelliptic type. Since the solution in the two planes of rotation is similar we will focus here on the case $y=-1$.

On the ``axis'' of $\psi$-rotation, we have $y=-1$, $\Psi =0$ and $p_y=\frac{\partial S}{\partial y}=0$. Then the Hamilton-Jacobi equation depends on the coordinate $x$ only
\begin{align}
 0 &= m^2-D^{2/3}(x,-1)\frac{H(x,-1)}{H(-1,x)}E^2 +D^{-1/3}(x,-1)\frac{(x+1)^2(1-\nu)^2}{R^2H(x,-1)}\left\lbrace G(x) \left(\frac{\partial S}{\partial x}\right) ^2\right. \nonumber \\
   & +\left. \frac{(\Phi+c\Omega_\phi E)^2}{(1-\nu)^2} \left[ \frac{\beta(x)}{G(x)}-\frac{\nu[2+\nu(1-\nu)+\lambda(2-3\nu)]}{1-\lambda+\nu}\right] \right\rbrace \, .
\end{align}

From this we get the derivative of the action $S$
\begin{align}
 \left(\frac{\partial S}{\partial x}\right) ^2 &=\frac{D^{1/3}(x,-1)R^2 H(x,-1)}{(x+1)^2(1-\nu)^2G(x)}\left( D^{2/3}(x,-1)\frac{H(x,-1)}{H(-1,x)}E^2-m^2 \right) \nonumber \\
 & - \frac{(\Phi+c\Omega_\phi E)^2}{(1-\nu)^2G(x)} \left( \frac{\beta(x)}{G(x)}-\frac{\nu[2+\nu(1-\nu)+\lambda(2-3\nu)]}{1-\lambda+\nu}\right) =: X_S \, . \nonumber \\
\end{align}
and the ansatz for the action $S$ in the Hamilton-Jacobi equation (see Section~\ref{sec:GeoMotion}) becomes
\begin{equation}
 S = \frac{1}{2}m^2\tau -Et + \Phi\phi + \int \! \sqrt{X_S} \, \mathrm{d}x \, .
\end{equation}

\textls[-25]{Following the Hamilton-Jacobi formalism we set the partial derivatives of $S$ with respect to the constants $m^2$, $E$ and $\Phi$ to zero, which gives us three differential equations of motion}
\begin{align}
 \frac{\mathrm{d}x}{\mathrm{d}\gamma} &= \sqrt{X(x)} \label{eqn:ring-x-equation}\\
 \frac{\mathrm{d}\phi}{\mathrm{d}\gamma} &= \frac{(x+1)H(x,-1)H(-1,x)}{(1+\nu-\lambda)^2G(x)}(\Phi +c\Omega_\phi E) \label{eqn:ring-phi-equation}\\
 \frac{\mathrm{d}t}{\mathrm{d}\gamma} &= R^2 E\frac{D(x,-1)H^2(x,-1)}{(x+1)H(-1,x)}- \frac{(x+1)H(x,-1)H(-1,x)}{(1+\nu-\lambda)^2G(x)}c\Omega_\phi(\Phi +c\Omega_\phi E) \label{eqn:ring-t-equation}
\end{align}
with
\begin{align}
 X(x) &= (1-\nu)^2 \frac{H(x,-1)}{H(-1,x)} \left\lbrace R^2 G(x) \left[D(x,-1) H(x,-1)E^2 - D^{1/3}(x,-1) H(-1,x)\delta\right] \right. \nonumber \\
 & \left. - \frac{(x+1)^2}{(1-\lambda +\nu)^2} \left[ H(-1,x)\Phi +cR\lambda \sqrt{\nu}\sqrt{2((1+\nu)^2-\lambda^2)}(1-x^2)E \right] ^2 \right\rbrace \nonumber \\
\end{align}
and
\begin{align}
 H(-1,x) &= (1-\lambda)^2-\nu ^2 + \nu x^2 (1-\lambda ^2-\nu ^2+2\lambda\nu) \nonumber\\
 H(x,-1) &= 1+\lambda ^2 -\nu ^2-2\lambda\nu (1-x^2)+2\lambda x (1-\nu ^2) + x^2\nu(1-\lambda ^2-\nu ^2) \nonumber\\
 D(x,-1) &= 1+\frac{s^2}{H(x,-1)}[2\lambda(1-\nu)(x+1)(1+\nu x)] \nonumber\\
 \Omega _\phi &= \frac{R\lambda\sqrt{2((1+\nu)^2-\lambda^2)}}{H(-1,x)}(1-x^2)\sqrt{\nu} \, .
\end{align}

We also defined $\gamma = \frac{(x+1)}{D(x,-1)^{1/3}R^2 H(x,-1)} \tau$ to simplify the equations of~motion.\\

It is possible to solve the equations of motion \eqref{eqn:ring-x-equation}--\eqref{eqn:ring-t-equation} if $X(x)$ is a polynomial, which happens in two~cases 
\begin{enumerate}
 \item $D(x,-1)=1$ (which implies $c=1$ and $s=0$): This case represents the motion of photons or particles around an uncharged doubly spinning black ring.
 \item $\delta=0$: In this case the motion of photons around a charged doubly spinning black ring is described.
\end{enumerate}

In both cases the equations of motion are of hyperelliptic type (genus $g=2$), since the polynomial $X = \sum_{i=1}^6 a_i x^i$ is of 6th~order.

The substitution $x=\pm \frac{1}{u}+x_Z$, where $x_Z$ is a zero of $X$, transforms $X$ into a polynomial of order five and the $x$-equation \eqref{eqn:ring-x-equation} becomes
\begin{equation}
 \left(u\frac{\dd u}{\dd \gamma}\right)^2=\sum_{i=0}^5 b_iu^i=:P_5(u) \, .
\end{equation}

A separation of variables yields the hyperelliptic integral
\begin{equation}
    \gamma - \gamma_{\mathrm in} = \int _{u_{\mathrm in}}^u \frac{u'\dd u'}{\sqrt{P_5(u')}}
\end{equation}

As described in Section~\ref{sec:hyperell}, see Equations \eqref{eqn:hyper-firstkind-solution} and \eqref{eqn:xinf-2}, the~solution of the above equation is
\begin{equation}
 u= -\frac{\sigma_1(\vec{\gamma}_{\infty})}{\sigma_2(\vec{\gamma}_{\infty})} \, .
\end{equation}

A resubstitution yields the full solution of \eqref{eqn:ring-x-equation}
\begin{equation}
 x(\gamma)= \mp \frac{\sigma_2(\vec{\gamma}_{\infty})}{\sigma_1(\vec{\gamma}_{\infty})}+x_Z \, .
\end{equation}
where $\sigma_i$ is the $i$th derivative of the $\sigma$-function and
\begin{equation}
	\vec{\gamma}_\infty = 
	\left( 
	\begin{array}{c}
		\gamma_1 \\
		\gamma-\gamma_\mathrm{in}''
	\end{array}
	\right) 
\end{equation}
with $\gamma_\mathrm{in}''= \gamma_\mathrm{in}+\int_{u_\mathrm{in}}^{\infty}\! \frac{u \, \dd u}{\sqrt{ P_5(u)}}$.  $\gamma_1$ is determined by the condition $\sigma (\vec{\gamma}_\infty)=0$.\\

Next we will solve the $\phi$-equation \eqref{eqn:ring-phi-equation} of the black ring. Using Equation \eqref{eqn:ring-x-equation} the $\phi$-equation can be written as
\begin{equation}
\phi - \phi_{\rm in} = \int _{x_{\rm in}}^x \! \frac{(x+1)H(x,-1)H(-1,x)}{(1+\nu-\lambda)^2G(x)}(\Phi +c\Omega_\phi E) \frac{\mathrm{d}x}{\sqrt{X(x)}} \, .
\end{equation}

The substitution $x=\pm \frac{1}{u}+x_Z$ and a partial fraction decomposition yields
\begin{equation}
 \phi - \phi_{\rm in} = \int _{u_{\rm in}}^u \! \left( \sum _{i=1}^3 \frac{K_i}{u-p_i} + K_4 + K_5 u \right) \,  \frac{\mathrm{d}u}{\sqrt{P_5(u)}}\, .
\end{equation}

The constants $K_i$ and the poles $p_i$ depend on the parameters of the black ring and the test~particle.

The hyperelliptic integrals of the first kind are known from the solution of the $x$-Equation \eqref{eqn:ring-x-equation} and thus
\begin{equation}
    \int _{u_{\rm in}}^u \! K_5 \,  \frac{u\mathrm{d}u}{\sqrt{P_5(u)}} = K_5 (\gamma-\gamma_{\rm in})\, .
\end{equation}
and
\begin{equation}
    \int _{u_{\rm in}}^u \!  K_4 \,  \frac{\mathrm{d}u}{\sqrt{P_5(u)}} =  K_4 \left(\gamma_1+\int_{y_\mathrm{in}}^\infty\!  \frac{\mathrm{d}u}{\sqrt{P_5(u)}}\right)
\end{equation}
where $\int_{y_\mathrm{in}}^\infty\!  \frac{\mathrm{d}u}{\sqrt{P_5(u)}}$ can be calculated in terms of the periods if $y_\mathrm{in}$ is chosen to be a zero of $P_5$.

The hyperelliptic integral of the third kind
\begin{equation}
    \int _{u_{\rm in}}^u \! \sum _{i=1}^3 \frac{K_i}{u-p_i} \,  \frac{\mathrm{d}u}{\sqrt{P_5(u)}}
\end{equation}
can be solved with the solution equation \eqref{eqn:sol-thirdkind} in Section~\ref{sec:hyperell}.

The complete solution of the $\phi$-equation \eqref{eqn:ring-phi-equation} is
\begin{align}
    \phi &= \sum _{i=1}^3 \left[\frac{2}{\sqrt{P_5{(p_i)}}}\int_{u_{\rm in}}^u{\dd\vec{z}^T}\int_{e_2}^{p_i}{\dd\vec{y}}
+\ln{\left(\frac{\sigma(\int_{\infty}^u{\dd\vec{z}}-\int_{e_2}^{p_i}{\dd\vec{z}})}
{\sigma(\int_{\infty}^u{\dd\vec{z}}+\int_{e_2}^{p_i}{\dd\vec{z}})}\right)}
-\ln{\left(\frac{\sigma(\int_{\infty}^{u_{\rm in}}{\dd\vec{z}}-\int_{e_2}^{p_i}{\dd\vec{z}})}
{\sigma(\int_{\infty}^{u_{\rm in}}{\dd\vec{z}}+\int_{e_2}^{p_i}{\dd\vec{z}})}\right)}\right]\nonumber\\
    & + K_4 \left(\gamma_1+\int_{y_\mathrm{in}}^\infty\!  \frac{\mathrm{d}u}{\sqrt{P_5(u)}}\right) + K_5 (\gamma-\gamma_{\rm in}) + \phi_{\rm in} \, .
\end{align}

Analogously, the~solution of the $t$-equation \eqref{eqn:ring-t-equation} can be~found.\\

\begin{figure}[H]
\begin{tabular}{ccc}

        {\includegraphics[width=0.31\linewidth]{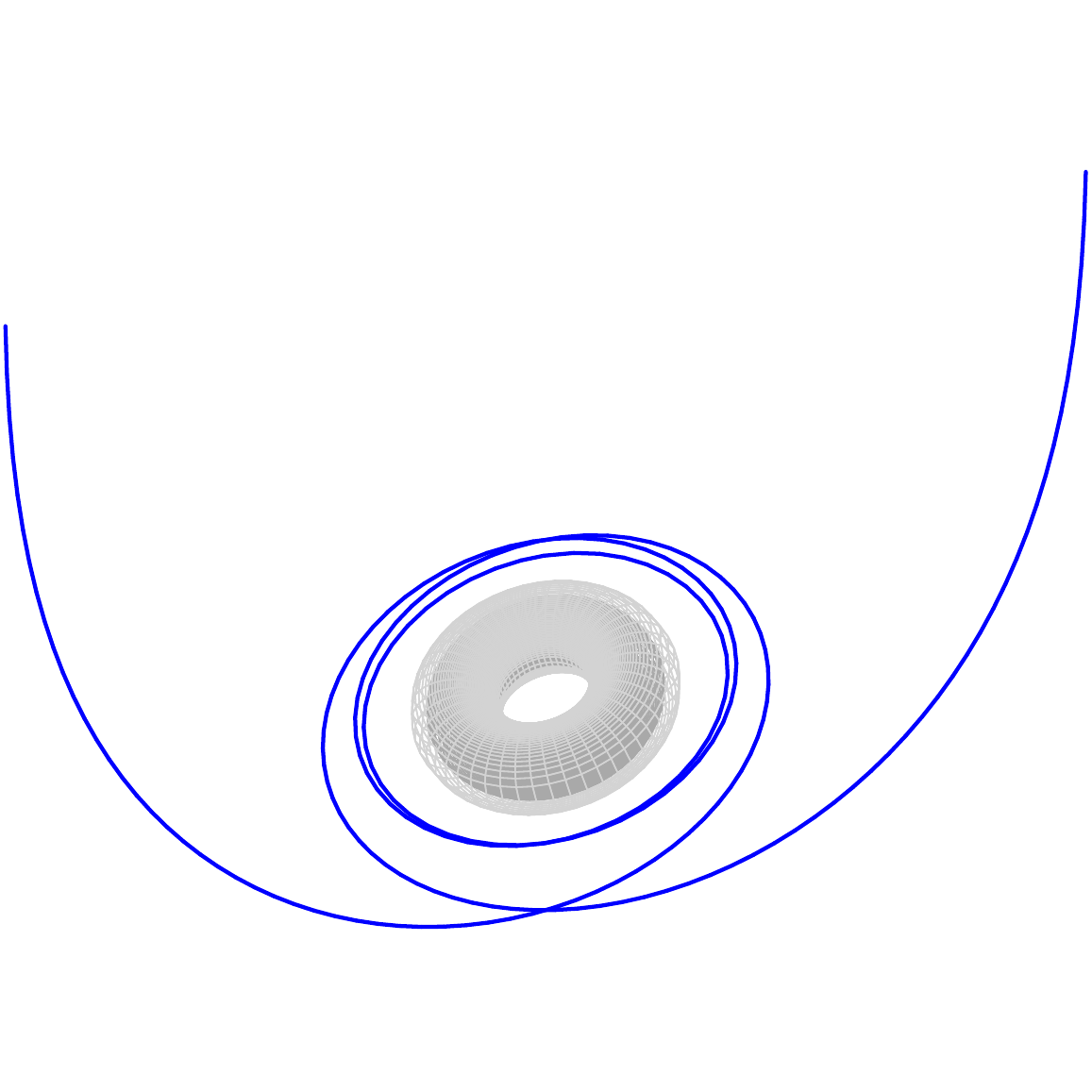}} &
  
        {\includegraphics[width=0.31\linewidth]{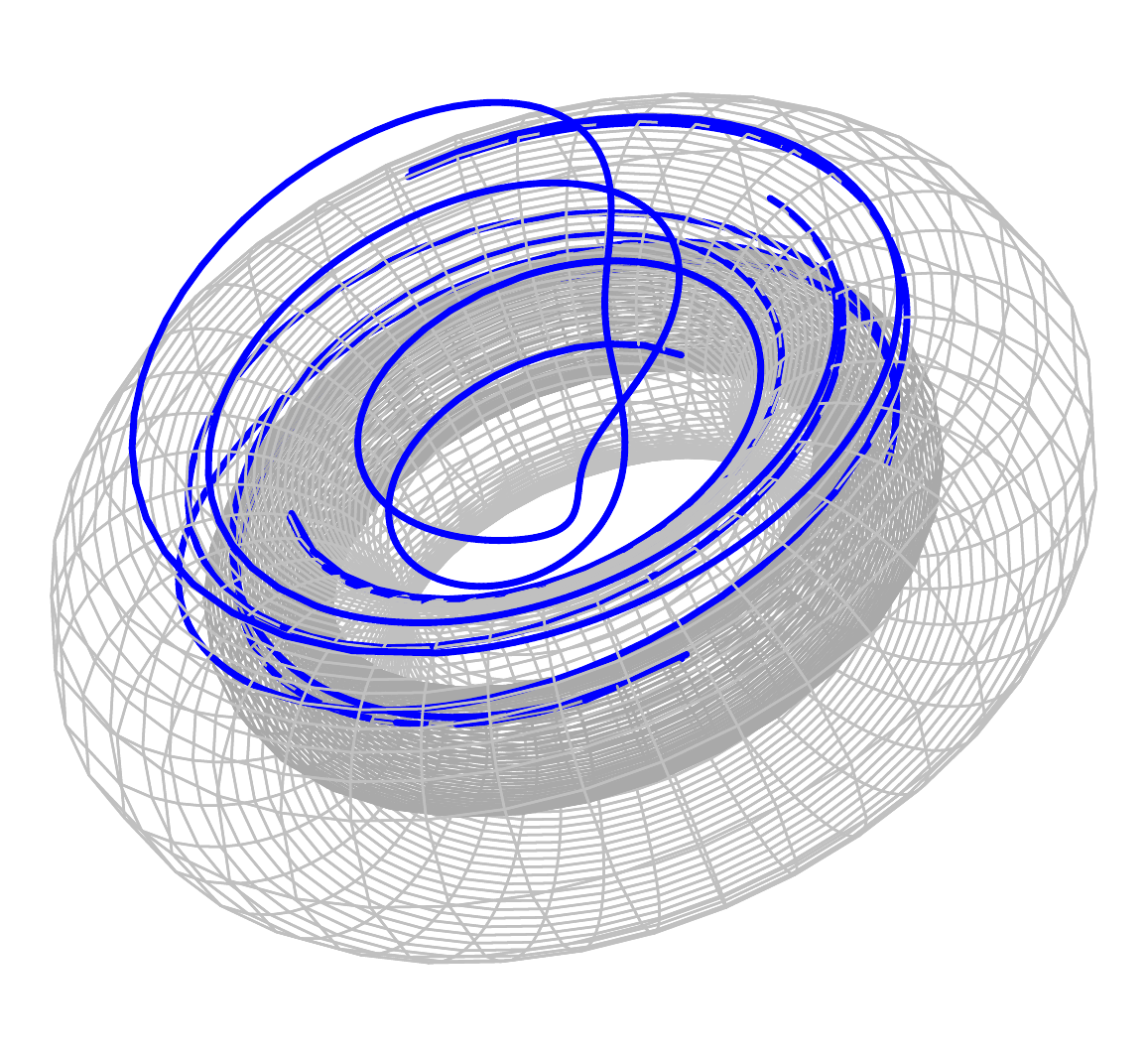}} &
  
        {\includegraphics[width=0.31\linewidth]{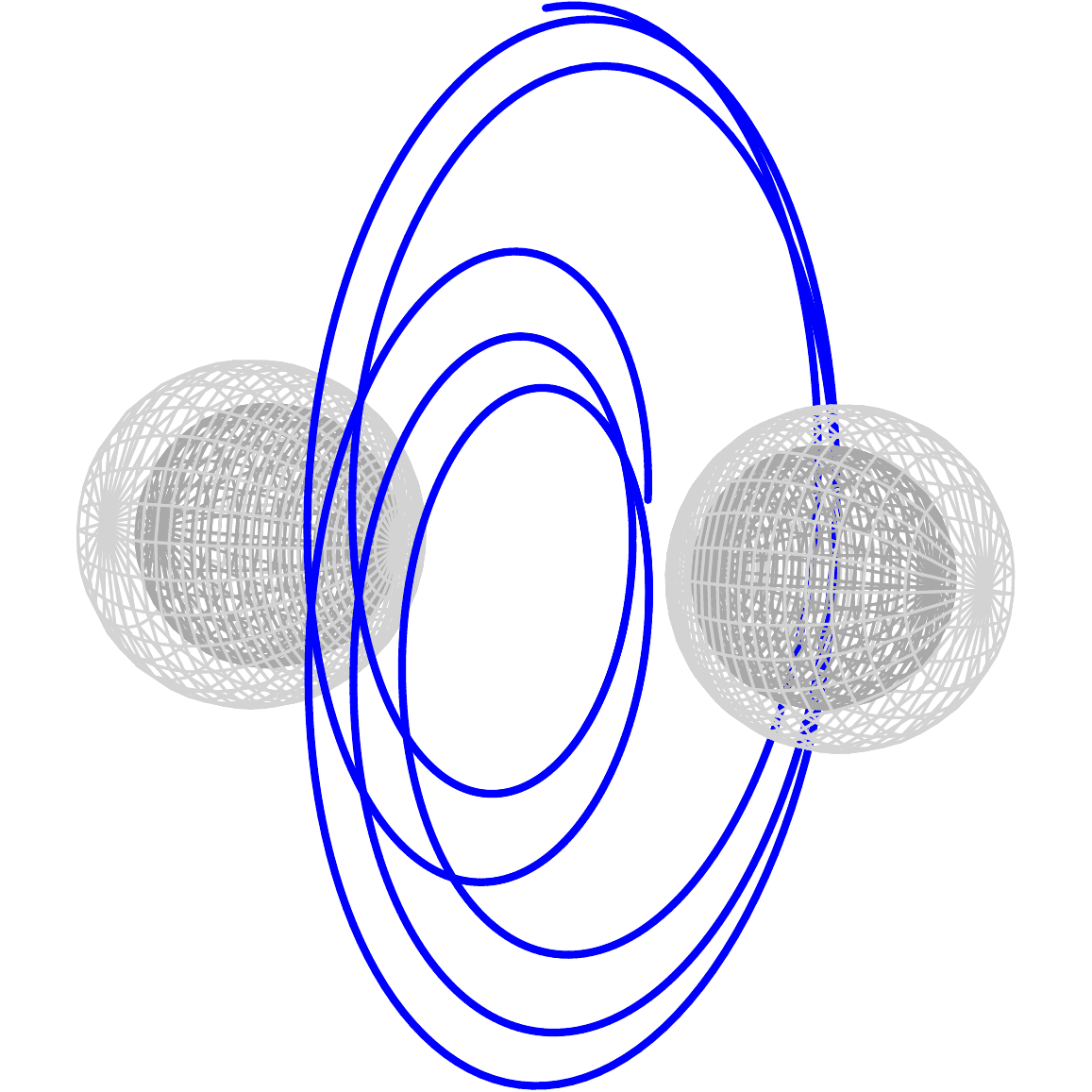}} \\
         ({\bf a})&({\bf b})& ({\bf c})\\
\end{tabular}
        
	\caption{Orbits of particles (blue curves) around the black ring. The~horizons are depicted as grey tori or~spheres. (\textbf{a}) Escape orbit in the equatorial plane ($x=\pm 1$); (\textbf{b}) Many-world bound orbit in the case $E=m=0$; (\textbf{c}) Bound orbit in the plane of $\psi$-rotation ($y=-1$). }
	\label{fig:ring-orbits}
\end{figure}

Using the analytical solutions we can plot orbits in the black ring spacetime. Figure~\ref{fig:ring-orbits}a shows an escape orbit in the equatorial plane ($x=\pm 1$) of the black ring. In~Figure~\ref{fig:ring-orbits}b a many-world bound orbit for light in the case $E=\delta=0$ is depicted. As~discussed above for Figure~\ref{fig:dyon-orbits}c, also for this ring spacetime the maximal analytic extension consists of an infinite set of \textit{worlds}, intermediate regions and regions with a singularity. A~bound orbit of a particle that crosses both outer and inner horizons twice then emerges into another world, only to enter the black ring again, and~repeat this whole process periodically.
A bound orbit is shown in Figure~\ref{fig:ring-orbits}c. In~the black ring spacetime bound orbits only exist in the plane of $\psi$-rotation ($y=-1$). Interestingly, in~higher dimensional spherical black hole spacetimes, such as the higher dimensional Schwarzschild and Myers-Perry black holes, stable bound orbits are not possible. In~the Myers-Perry spacetime stable bound orbits can only be found hidden behind the horizons. Therefore, the~bound orbit in Figure~\ref{fig:ring-orbits}c seems to be a particular feature of the black ring. We note, that Figure~\ref{fig:ring-orbits} illustrates orbits in the black ring spacetime by making use of different projections. Figure~\ref{fig:ring-orbits}a,b suppress one spatial coordinate of the $S^2$ and retain the $S^1$, whereas Figure~\ref{fig:ring-orbits}c suppresses the $S^1$ while retaining the $S^2$. Therefore the horizons look connected and ringlike in (a) and (b), whereas in (c) the horizon appears as two separate~spheres.

\unskip

\section{Conclusions}

Geodesic motion in black hole spacetimes is of utmost relevance for fundamental physics and astrophysics, as~well as for technological applications. 
Since exact solutions of the differential equations provide arbitrary accuracy, they are the means of choice.
Numerous black hole spacetimes allow for exact solutions, based on elliptic and hyperelliptic integrals.
Whereas the elliptic case has been widely studied, hyperelliptic geodesic equations have received much less attention~\cite{Kraniotis:2004cz,Hackmann:2008zz,Hackmann:2008tu,Hackmann:2009nh,Hackmann:2010zz,Enolski:2010if,Enolski:2011id,Garcia:2013zud,Hackmann:2014tga,Hendi:2020yah, Grunau:2019bsd, Flathmann:2016knq, Soroushfar:2016esy,Hoseini:2016nzw, Flathmann:2019mlj, Chatterjee:2019rym}.

Here we have reviewed the general method for constructing solutions of hyperelliptic geodesic equations~\cite{Enolski:2010if}, and~we have illustrated the method for the $g=2$ case with several examples: 9- and 11-dimensional Schwarzschild black holes, 4-dimensional supergravity black holes, and~5-dimensional black rings.
Whereas numerous interesting spacetimes with $g=2$ hyperelliptic geodesic equations are still awaiting analysis, this is even more so for spacetimes with $g>2$ hyperelliptic equations~\cite{Enolski:2010if,Enolski:2011id}.

However, in~alternative theories of gravity also geodesic equations can arise, that are characterized by polynomials of the more general type
\begin{equation}
y^n + P^{(n-2)}_{m_{n-2}} (x) y^{n-2} + P^{(n-3)}_{m_{n-3}} (x) y^{n-3} + \ldots + P^{(1)}_{m_1}(x) y + P^{(0)}_{m_0}(x) = 0 \, ,
\end{equation}
{with $P^{(n)}_m(x)$ polynomials of order $m$ in $x$.
For instance, in~Ho\v{r}ava--Lifshitz black hole spacetimes {\cite{Kehagias:2009is, Lu:2009em, Park:2009zra}} as well as Gau{\ss}--Bonnet black hole~{\cite{Boulware:1985wk}} spacetimes quartic equations of the form}
\begin{equation}
y^4 + P_m^{(2)}(x) y^2 + P_n^{(0)}(x) = 0 \,  \label{prym}
\end{equation}
arise. 
While in special cases such curves can be reduced to lower genera and the above methods become applicable~\cite{Enolski:2011id}, for~the general set of geodesic equations so far only numerical analysis has been performed~\cite{Enolskii:2011rk}. 
The extension of the above methods to obtain exact solutions also in such general cases remains a challenge to be~tackled.

\vspace{6pt} 

\authorcontributions{S.G. has contributed the major part of the text and all of the figures, J.K. has contributed the remaining part of the~text. All authors have read and agreed to the published version of the manuscript.}

\funding{This work was supported by the DFG Research Training Group 1620 {\sl Models of~Gravity}.}

\dataavailability{Data is available by request from the authors.} 

\acknowledgments{We are deeply grateful to our numerous collaborators on this subject over the years and, in~particular, to~the late Victor Enolski, to~Valeria Diemer (n\'ee Kagramanova), Eva Hackmann, and~Claus L\"ammerzahl, who introduced us to this interesting topic.}

\conflictsofinterest{The authors declare no conflict of~interest.} 


%


\begin{adjustwidth}{-\extralength}{0cm}

\reftitle{References}


\begin{thebibliography}{999}

\bibitem[Ashby(2003)]{Ashby:2003}
Ashby, N.
\newblock {Relativity in the Global Positioning System}.
\newblock {\em Living Rev. Relativ.} {\bf 2003}, {\em 6}, 1--42; doi:10.12942/lrr-2003-1.


\bibitem[Will(2018)]{Will:2018bme}
Will, C.M.
\newblock {\em {Theory and Experiment in Gravitational Physics}}; Cambridge
  University Press: Cambridge, UK,  2018.

\bibitem[Oppenheimer and Snyder(1939)]{Oppenheimer:1939ue}
Oppenheimer, J.R.; Snyder, H.
\newblock {On Continued gravitational contraction}.
\newblock {\em Phys. Rev.} {\bf 1939}, {\em 56},~455--459, doi:10.1103/PhysRev.56.455.

\bibitem[Penrose(1965)]{Penrose:1964wq}
Penrose, R.
\newblock {Gravitational collapse and space-time singularities}.
\newblock {\em Phys. Rev. Lett.} {\bf 1965}, {\em 14},~57--59, doi:10.1103/PhysRevLett.14.57.

\bibitem[Penrose(1969)]{Penrose:1969pc}
Penrose, R.
\newblock {Gravitational collapse: The role of general relativity}.
\newblock {\em Riv. Nuovo Cim.} {\bf 1969}, {\em 1},~252--276.

\bibitem[Penrose(2002)]{Penrose:1969pc-2}
Penrose, R.
\newblock {{''Golden Oldie'': Gravitational collapse: The role of general
  relativity}}.
\newblock {\em Gen. Rel. Grav.} {\bf 2002}, {\em 34},~1141--1165, doi:10.1023/A:1016578408204.

\bibitem[Webster and Murdin(1972)]{Webster:1972bsw}
Webster, B.L.; Murdin, P.
\newblock {Cygnus X-1-a Spectroscopic Binary with a Heavy Companion?}
\newblock {\em Nature} {\bf 1972}, {\em 235},~37--38, doi:10.1038/235037a0.

\bibitem[Bolton(1972)]{Bolton:1972bun}
Bolton, C.T.
\newblock {Dimensions of the Binary System HDE 226868 = Cygnus X-1}.
\newblock {\em Nat. Phys. Sci.} {\bf 1972}, {\em 240},~124--127, doi:10.1038/physci240124a0.

\bibitem[Kormendy and Richstone(1995)]{Kormendy:1995er}
Kormendy, J.; Richstone, D.
\newblock {Inward bound: The Search for supermassive black holes in galactic
  nuclei}.
\newblock {\em Ann. Rev. Astron. Astrophys.} {\bf 1995}, {\em 33},~581, doi:10.1146/annurev.aa.33.090195.003053.

\bibitem[Eckart and Genzel(1996)]{Eckart:1996zz}
Eckart, A.; Genzel, R.
\newblock {Observations of stellar proper motions near the Galactic Centre}.
\newblock {\em Nature} {\bf 1996}, {\em 383},~415--417, doi:10.1038/383415a0.

\bibitem[Ghez \em{et~al.}(1998)Ghez, Klein, Morris, and Becklin]{Ghez:1998ph}
Ghez, A.M.; Klein, B.L.; Morris, M.; Becklin, E.E.
\newblock {High proper motion stars in the vicinity of Sgr A*: Evidence for a
  supermassive black hole at the center of our galaxy}.
\newblock {\em Astrophys. J.} {\bf 1998}, {\em 509},~678--686, doi:10.1086/306528.

\bibitem[Celotti \em{et~al.}(1999)Celotti, Miller, and Sciama]{Celotti:1999tg}
Celotti, A.; Miller, J.C.; Sciama, D.W.
\newblock {Astrophysical evidence for the existence of black holes: Topical
  review}.
\newblock {\em Class. Quant. Grav.} {\bf 1999}, {\em 16},~A3, doi:10.1088/0264-9381/16/12A/301.

\bibitem[Ferrarese and Ford(2005)]{Ferrarese:2004qr}
Ferrarese, L.; Ford, H.
\newblock {Supermassive black holes in galactic nuclei: Past, present and
  future research}.
\newblock {\em Space Sci. Rev.} {\bf 2005}, {\em 116},~523--624, doi:10.1007/s11214-005-3947-6.

\bibitem[Abbott \em{et~al.}(2016)Abbott et~al.]{LIGOScientific:2016aoc}
Abbott, B.P.; Abbott, R.; Abbott, T.D.; Abernathy, M.R.; Acernese, F.; Ackley, K.; Adams, C.; Adams, T.; Addesso, P.; Adhikari, R. X.; et al.
\newblock {Observation of Gravitational Waves from a Binary Black Hole Merger}.
\newblock {\em Phys. Rev. Lett.} {\bf 2016}, {\em 116},~061102, doi:10.1103/PhysRevLett.116.061102.

\bibitem[Akiyama \em{et~al.}(2019)Akiyama
  et~al.]{EventHorizonTelescope:2019dse}
Akiyama, K.; Alberdi, A.; Alef, W.; Asada, K.; Azulay, R.; Baczko, An.; Ball, D.; Baloković, M.; Barrett, J.; Bintley, D.; et al.
\newblock {First M87 Event Horizon Telescope Results. I. The Shadow of the
  Supermassive Black Hole}.
\newblock {\em Astrophys. J. Lett.} {\bf 2019}, {\em 875},~L1, doi:10.3847/2041-8213/ab0ec7.

\bibitem[Hagihara(1931)]{Hagihara:1931}
Hagihara, Y.
\newblock {Theory of the Relativistic Trajectories in a Gravitational Field of
  Schwarzschild}.
\newblock {\em Jpn. J. Astron. Geophys.} {\bf 1931}, {\em 8},~67.

\bibitem[Carter(1968)]{Carter:1968rr}
Carter, B.
\newblock {Global structure of the Kerr family of gravitational fields}.
\newblock {\em Phys. Rev.} {\bf 1968}, {\em 174},~1559--1571, doi:10.1103/PhysRev.174.1559.

\bibitem[Bardeen(1973)]{Bardeen:1973}
Bardeen, J.
\newblock Timelike and null geodesics in the {K}err metric. In {\em Black
  Holes}; DeWitt, C., DeWitt, B., Eds.; Gordon and Breach: New York,  NY, USA, 1973; p.
  215.

\bibitem[Chandrasekhar(1985)]{Chandrasekhar:1985kt}
Chandrasekhar, S.
\newblock {\em {The Mathematical Theory of Black Holes}}; Clarendon Press: Oxford, UK,  1985.

\bibitem[Perlick(2004)]{Perlick:2004tq}
Perlick, V.
\newblock {Gravitational lensing from a spacetime perspective}.
\newblock {\em Living Rev. Rel.} {\bf 2004}, {\em 7},~9.

\bibitem[Kraniotis(2007)]{Kraniotis:2007zz}
Kraniotis, G.V.
\newblock {Periapsis and gravitomagnetic precessions of stellar orbits in Kerr
  and Kerr-de Sitter black hole spacetimes}.
\newblock {\em Class. Quant. Grav.} {\bf 2007}, {\em 24},~1775--1808, doi:10.1088/0264-9381/24/7/007.

\bibitem[Kagramanova \em{et~al.}(2010)Kagramanova, Kunz, Hackmann, and
  L\"ammerzahl]{Kagramanova:2010bk}
Kagramanova, V.; Kunz, J.; Hackmann, E.; L\"ammerzahl, C.
\newblock {Analytic treatment of complete and incomplete geodesics in Taub-NUT
  space-times}.
\newblock {\em Phys. Rev. D} {\bf 2010}, {\em 81},~124044, doi:10.1103/PhysRevD.81.124044.

\bibitem[Grunau and Kagramanova(2011)]{Grunau:2010gd}
Grunau, S.; Kagramanova, V.
\newblock {Geodesics of electrically and magnetically charged test particles in
  the Reissner-Nordstr\"om space-time: Analytical solutions}.
\newblock {\em Phys. Rev. D} {\bf 2011}, {\em 83},~044009, doi:10.1103/PhysRevD.83.044009.

\bibitem[Hackmann \em{et~al.}(2010)Hackmann, Hartmann, L\"ammerzahl, and
  Sirimachan]{Hackmann:2010ir}
Hackmann, E.; Hartmann, B.; L\"ammerzahl, C.; Sirimachan, P.
\newblock {Test particle motion in the space-time of a Kerr black hole pierced
  by a cosmic string}.
\newblock {\em Phys. Rev. D} {\bf 2010}, {\em 82},~044024, doi:10.1103/PhysRevD.82.044024.

\bibitem[Kraniotis(2011)]{Kraniotis:2010gx}
Kraniotis, G.V.
\newblock {Precise analytic treatment of Kerr and Kerr-(anti) de Sitter black
  holes as gravitational lenses}.
\newblock {\em Class. Quant. Grav.} {\bf 2011}, {\em 28},~085021, doi:10.1088/0264-9381/28/8/085021.

\bibitem[Kagramanova and Reimers(2012)]{Kagramanova:2012hw}
Kagramanova, V.; Reimers, S.
\newblock {Analytic treatment of geodesics in five-dimensional Myers-Perry
  space--times}.
\newblock {\em Phys. Rev. D} {\bf 2012}, {\em 86},~084029, doi:10.1103/PhysRevD.86.084029.

\bibitem[Hackmann and Xu(2013)]{Hackmann:2013pva}
Hackmann, E.; Xu, H.
\newblock {Charged particle motion in Kerr-Newmann space-times}.
\newblock {\em Phys. Rev. D} {\bf 2013}, {\em 87},~124030, doi:10.1103/PhysRevD.87.124030.

\bibitem[Diemer and Smolarek(2013)]{Diemer:2013hgn}
Diemer, V.; Smolarek, E.
\newblock {Dynamics of test particles in thin-shell wormhole spacetimes}.
\newblock {\em Class. Quant. Grav.} {\bf 2013}, {\em 30},~175014, doi:10.1088/0264-9381/30/17/175014.

\bibitem[Diemer and Kunz(2014)]{Diemer:2013fza}
Diemer, V.; Kunz, J.
\newblock {Supersymmetric rotating black hole spacetime tested by geodesics}.
\newblock {\em Phys. Rev. D} {\bf 2014}, {\em 89},~084001, doi:10.1103/PhysRevD.89.084001.

\bibitem[Grunau and Khamesra(2013)]{Grunau:2013oca}
Grunau, S.; Khamesra, B.
\newblock {Geodesic motion in the (rotating) black string spacetime}.
\newblock {\em Phys. Rev. D} {\bf 2013}, {\em 87},~124019, doi:10.1103/PhysRevD.87.124019.

\bibitem[Grenzebach \em{et~al.}(2014)Grenzebach, Perlick, and
  L\"ammerzahl]{Grenzebach:2014fha}
Grenzebach, A.; Perlick, V.; L\"ammerzahl, C.
\newblock {Photon Regions and Shadows of Kerr-Newman-NUT Black Holes with a
  Cosmological Constant}.
\newblock {\em Phys. Rev. D} {\bf 2014}, {\em 89},~124004, doi:10.1103/PhysRevD.89.124004.

\bibitem[Diemer \em{et~al.}(2014)Diemer, Kunz, L\"ammerzahl, and
  Reimers]{Diemer:2014lba}
Diemer, V.; Kunz, J.; L\"ammerzahl, C.; Reimers, S.
\newblock {Dynamics of test particles in the general five-dimensional
  Myers-Perry spacetime}.
\newblock {\em Phys. Rev. D} {\bf 2014}, {\em 89},~124026, doi:10.1103/PhysRevD.89.124026.

\bibitem[Flathmann and Grunau(2015)]{Flathmann:2015xia}
Flathmann, K.; Grunau, S.
\newblock {Analytic solutions of the geodesic equation for
  Einstein-Maxwell-dilaton-axion black holes}.
\newblock {\em Phys. Rev. D} {\bf 2015}, {\em 92},~104027, doi:10.1103/PhysRevD.92.104027.

\bibitem[Kraniotis(2014)]{Kraniotis:2014paa}
Kraniotis, G.V.
\newblock {Gravitational lensing and frame dragging of light in the Kerr-Newman 
  and the Kerr-Newman-(anti) de Sitter black hole spacetimes}.
\newblock {\em Gen. Rel. Grav.} {\bf 2014}, {\em 46},~1818, doi:10.1007/s10714-014-1818-8.

\bibitem[Kraniotis(2015)]{Kraniotis:2015kfd}
Kraniotis, G.
\newblock {Gravitational lensing and frame dragging of light in the Kerr-Newman
  and the Kerr-Newman-(anti) de Sitter black hole spacetimes}.
\newblock {\em Proc. Sci}. {\bf 2015}, {\em PLANCK2015},~073, doi:10.22323/1.258.0073.

\bibitem[Paranjape and Reimers(2016)]{Paranjape:2016oly}
Paranjape, S.; Reimers, S.
\newblock {Dynamics of test particles in the five-dimensional, charged,
  rotating Einstein-Maxwell-Chern-Simons spacetime}.
\newblock {\em Phys. Rev. D} {\bf 2016}, {\em 94},~124003, doi:10.1103/PhysRevD.94.124003.

\bibitem[Grunau \em{et~al.}(2018)Grunau, Neumann, and Reimers]{Grunau:2017uzf}
Grunau, S.; Neumann, H.; Reimers, S.
\newblock {Geodesic motion in the five-dimensional Myers-Perry-AdS spacetime}.
\newblock {\em Phys. Rev. D} {\bf 2018}, {\em 97},~044011, doi:10.1103/PhysRevD.97.044011.

\bibitem[Eickhoff and Reimers(2018)]{Eickhoff:2018msh}
Eickhoff, K.; Reimers, S.
\newblock {Dynamics of test particles in the five-dimensional G\"odel
  spacetime}.
\newblock {\em Phys. Rev. D} {\bf 2018}, {\em 98},~044050, doi:10.1103/PhysRevD.98.044050.

\bibitem[Willenborg \em{et~al.}(2018)Willenborg, Grunau, Kleihaus, and
  Kunz]{Willenborg:2018zsv}
Willenborg, F.; Grunau, S.; Kleihaus, B.; Kunz, J.
\newblock {Geodesic motion around traversable wormholes supported by a massless
  conformally-coupled scalar field}.
\newblock {\em Phys. Rev. D} {\bf 2018}, {\em 97},~124002, doi:10.1103/PhysRevD.97.124002.

\bibitem[Drawer and Grunau(2020)]{Drawer:2020mpw}
Drawer, J.C.; Grunau, S.
\newblock {Geodesic motion around a supersymmetric $\hbox {AdS}_5$ black hole}.
\newblock {\em Eur. Phys. J. C} {\bf 2020}, {\em 80},~536, doi:10.1140/epjc/s10052-020-8101-9.

\bibitem[Baker(1995)]{Baker:1995}
\textls[-25]{Baker, H.M.
\newblock {\em {Abelian Functions: Abel's Theorem and the Allied Theory of
  Theta Functions}}; Cambridge University Press: Cambridge, UK, 1995.}

\bibitem[Kraniotis and Whitehouse(2003)]{Kraniotis:2003ig}
Kraniotis, G.V.; Whitehouse, S.B.
\newblock {Exact calculation of the perihelion precession of mercury in general
  relativity, the cosmological constant and jacobi's inversion problem}.
\newblock {\em Class. Quant. Grav.} {\bf 2003}, {\em 20},~4817--4835, doi:10.1088/0264-9381/20/22/007.

\bibitem[Kraniotis(2004)]{Kraniotis:2004cz}
Kraniotis, G.V.
\newblock {Precise relativistic orbits in Kerr and Kerr–(anti) de Sitter
  spacetimes}.
\newblock {\em Class. Quant. Grav.} {\bf 2004}, {\em 21},~4743--4769, doi:10.1088/0264-9381/21/19/016.

\bibitem[Kraniotis(2005)]{Kraniotis:2005zm}
Kraniotis, G.V.
\newblock {Frame-dragging and bending of light in Kerr and Kerr-(anti) de
  Sitter spacetimes}.
\newblock {\em Class. Quant. Grav.} {\bf 2005}, {\em 22},~4391--4424, doi:10.1088/0264-9381/22/21/001.

\bibitem[Enolski \em{et~al.}(2003)Enolski, Pronine, and Richter]{Enolski:2003}
Enolski, V.; Pronine, M.; Richter, P.H.
\newblock {Double Pendulum and $\theta$-Divisor}.
\newblock {\em J. Nonlinear Sci.} {\bf 2003}, {\em 13},~157--174, doi:10.1007/s00332-002-0514-0.

\bibitem[Hackmann and L\"ammerzahl(2008)]{Hackmann:2008zza}
Hackmann, E.; L\"ammerzahl, C.
\newblock {Complete Analytic Solution of the Geodesic Equation in
  Schwarzschild- (Anti-) de Sitter Spacetimes}.
\newblock {\em Phys. Rev. Lett.} {\bf 2008}, {\em 100},~171101, doi:10.1103/PhysRevLett.100.171101.

\bibitem[Hackmann and Lämmerzahl(2008)]{Hackmann:2008zz}
Hackmann, E.; Lämmerzahl, C.
\newblock {Geodesic equation in Schwarzschild- (anti-) de Sitter space-times:
  Analytical solutions and applications}.
\newblock {\em Phys. Rev. D} {\bf 2008}, {\em 78},~024035, doi:10.1103/PhysRevD.78.024035.

\bibitem[Hackmann \em{et~al.}(2008)Hackmann, Kagramanova, Kunz, and
  Lämmerzahl]{Hackmann:2008tu}
Hackmann, E.; Kagramanova, V.; Kunz, J.; Lämmerzahl, C.
\newblock {Analytic solutions of the geodesic equation in higher dimensional
  static spherically symmetric space-times}.
\newblock {\em Phys. Rev. D} {\bf 2008}, {\em 78},~124018, 
  doi:{\changeurlcolor{black}\href{https://doi.org/10.1103/PhysRevD.78.124018}{\detokenize{10.1103/PhysRevD.78.124018}}}.

\bibitem[Hackmann \em{et~al.}(2009)Hackmann, Kagramanova, Kunz, and
  L\"ammerzahl]{Hackmann:2009nh}
Hackmann, E.; Kagramanova, V.; Kunz, J.; L\"ammerzahl, C.
\newblock {Analytic solutions of the geodesic equation in axially symmetric
  space-times}.
\newblock {\em EPL} {\bf 2009}, {\em 88},~30008, doi:10.1209/0295-5075/88/30008.

\bibitem[Hackmann \em{et~al.}(2010)Hackmann, L\"ammerzahl, Kagramanova, and
  Kunz]{Hackmann:2010zz}
Hackmann, E.; L\"ammerzahl, C.; Kagramanova, V.; Kunz, J.
\newblock {Analytical solution of the geodesic equation in Kerr-(anti) de
  Sitter space-times}.
\newblock {\em Phys. Rev. D} {\bf 2010}, {\em 81},~044020, doi:10.1103/PhysRevD.81.044020.

\bibitem[Enolski \em{et~al.}(2011)Enolski, Hackmann, Kagramanova, Kunz, and
  Lämmerzahl]{Enolski:2010if}
Enolski, V.; Hackmann, E.; Kagramanova, V.; Kunz, J.; Lämmerzahl, C.
\newblock {Inversion of hyperelliptic integrals of arbitrary genus with
  application to particle motion in General Relativity}.
\newblock {\em J. Geom. Phys.} {\bf 2011}, {\em 61},~899--921, doi:10.1016/j.geomphys.2011.01.001.

\bibitem[Enolski \em{et~al.}(2012)Enolski, Hartmann, Kagramanova, Kunz,
  Lämmerzahl, and Sirimachan]{Enolski:2011id}
Enolski, V.; Hartmann, B.; Kagramanova, V.; Kunz, J.; Lämmerzahl, C.;
  Sirimachan, P.
\newblock {Inversion of a general hyperelliptic integral and particle motion in
  Hořava–Lifshitz black hole space-times}.
\newblock {\em J. Math. Phys.} {\bf 2012}, {\em 53},~012504, doi:10.1063/1.3677831.

\bibitem[Walker and Penrose(1970)]{Walker:1970un}
Walker, M.; Penrose, R.
\newblock {On quadratic first integrals of the geodesic equations for type [22]
  spacetimes}.
\newblock {\em Commun. Math. Phys.} {\bf 1970}, {\em 18},~265--274, doi:10.1007/BF01649445.

\bibitem[Tangherlini(1963)]{Tangherlini:1963bw}
Tangherlini, F.R.
\newblock {Schwarzschild field in n dimensions and the dimensionality of space
  problem}.
\newblock {\em Nuovo Cim.} {\bf 1963}, {\em 27},~636--651, doi:10.1007/BF02784569.

\bibitem[Myers and Perry(1986)]{Myers:1986un}
Myers, R.C.; Perry, M.J.
\newblock {Black Holes in Higher Dimensional Space-Times}.
\newblock {\em Ann. Phys.} {\bf 1986}, {\em 172},~304, doi:10.1016/0003-4916(86)90186-7.

\bibitem[Frolov and Stojkovic(2003{\natexlab{a}})]{Frolov:2002xf}
Frolov, V.P.; Stojkovic, D.
\newblock {Quantum radiation from a five-dimensional rotating black hole}.
\newblock {\em Phys. Rev. D} {\bf 2003}, {\em 67},~084004, doi:10.1103/PhysRevD.67.084004.

\bibitem[Frolov and Stojkovic(2003{\natexlab{b}})]{Frolov:2003en}
Frolov, V.P.; Stojkovic, D.
\newblock {Particle and light motion in a space-time of a five-dimensional
  rotating black hole}.
\newblock {\em Phys. Rev. D} {\bf 2003}, {\em 68},~064011, doi:10.1103/PhysRevD.68.064011.

\bibitem[Page \em{et~al.}(2007)Page, Kubiznak, Vasudevan, and
  Krtous]{Page:2006ka}
Page, D.N.; Kubiznak, D.; Vasudevan, M.; Krtous, P.
\newblock {Complete integrability of geodesic motion in general Kerr-NUT-AdS
  spacetimes}.
\newblock {\em Phys. Rev. Lett.} {\bf 2007}, {\em 98},~061102, doi:10.1103/PhysRevLett.98.061102.

\bibitem[Kubiznak and Frolov(2007)]{Kubiznak:2006kt}
Kubiznak, D.; Frolov, V.P.
\newblock {Hidden Symmetry of Higher Dimensional Kerr-NUT-AdS Spacetimes}.
\newblock {\em Class. Quant. Grav.} {\bf 2007}, {\em 24},~F1--F6, doi:10.1088/0264-9381/24/3/F01.

\bibitem[Matsutani and Previato(2008)]{Matsutani:2008}
Matsutani, S.; Previato, E.
\newblock {Jacobi inversion on strata of the Jacobian of the $C_{rs}$ curve
  $y^r = f(x)$}.
\newblock {\em J. Math. Soc. Jpn.} {\bf 2008}, {\em
  60},~1009--1044, doi:10.2969/jmsj/06041009.

\bibitem[Ônishi(1998)]{Onishi:1998}
Ônishi, Y.
\newblock {Complex Multiplication Formulae for Hyperelliptic Curves of Genus
  Three}.
\newblock {\em Tokyo J. Math.} {\bf 1998}, {\em 21},~381--431, doi:10.3836/tjm/1270041822.

\bibitem[Mumford(1983)]{Mumford:1983}
Mumford, D.
\newblock {\em {Tata Lectures on Theta, Vol. I and II}}; Birkhäuser, Boston, MA, USA, 
  1983.

\bibitem[Chow and Comp\`ere(2014)]{Chow:2013gba}
Chow, D.D.K.; Comp\`ere, G.
\newblock {Dyonic AdS black holes in maximal gauged supergravity}.
\newblock {\em Phys. Rev. D} {\bf 2014}, {\em 89},~065003, doi:10.1103/PhysRevD.89.065003.

\bibitem[Flathmann and Grunau(2016)]{Flathmann:2016knq}
Flathmann, K.; Grunau, S.
\newblock {Analytic solutions of the geodesic equation for $U(1)^2$ dyonic
  rotating black holes}.
\newblock {\em Phys. Rev. D} {\bf 2016}, {\em 94},~124013, doi:10.1103/PhysRevD.94.124013.

\bibitem[Emparan and Reall(2002)]{Emparan:2001wn}
Emparan, R.; Reall, H.S.
\newblock {A Rotating black ring solution in five-dimensions}.
\newblock {\em Phys. Rev. Lett.} {\bf 2002}, {\em 88},~101101, doi:10.1103/PhysRevLett.88.101101.

\bibitem[Pomeransky and Sen'kov(2006)]{Pomeransky:2006bd}
Pomeransky, A.A.; Sen'kov, R.A.
\newblock {Black ring with two angular momenta.} \emph{arXiv} {\bf 2006}, arXiv:hep-th/0612005.

\bibitem[Elvang(2003)]{Elvang:2003yy}
Elvang, H.
\newblock {A Charged rotating black ring}.
\newblock {\em Phys. Rev. D} {\bf 2003}, {\em 68},~124016, doi:10.1103/PhysRevD.68.124016.

\bibitem[Hoskisson(2009)]{Hoskisson:2008qq}
Hoskisson, J.
\newblock {A Charged Doubly Spinning Black Ring}.
\newblock {\em Phys. Rev. D} {\bf 2009}, {\em 79},~104022, doi:10.1103/PhysRevD.79.104022.

\bibitem[Grunau \em{et~al.}(2012)Grunau, Kagramanova, Kunz, and
  L\"ammerzahl]{Grunau:2012ai}
Grunau, S.; Kagramanova, V.; Kunz, J.; L\"ammerzahl, C.
\newblock {Geodesic Motion in the Singly Spinning Black Ring Spacetime}.
\newblock {\em Phys. Rev. D} {\bf 2012}, {\em 86},~104002, doi:10.1103/PhysRevD.86.104002.

\bibitem[Grunau \em{et~al.}(2013)Grunau, Kagramanova, and Kunz]{Grunau:2012ri}
Grunau, S.; Kagramanova, V.; Kunz, J.
\newblock {Geodesic Motion in the (Charged) Doubly Spinning Black Ring
  Spacetime}.
\newblock {\em Phys. Rev. D} {\bf 2013}, {\em 87},~044054, doi:10.1103/PhysRevD.87.044054.

\bibitem[Garc\'\i{}a \em{et~al.}(2015)Garc\'\i{}a, Hackmann, Kunz,
  L\"ammerzahl, and Mac\'\i{}as]{Garcia:2013zud}
Garc\'\i{}a, A.; Hackmann, E.; Kunz, J.; L\"ammerzahl, C.; Mac\'\i{}as, A.
\newblock {Motion of test particles in a regular black hole
  space\textendash{}time}.
\newblock {\em J. Math. Phys.} {\bf 2015}, {\em 56},~032501, doi:10.1063/1.4913882.

\bibitem[Hackmann \em{et~al.}(2014)Hackmann, L\"ammerzahl, Obukhov, Puetzfeld,
  and Schaffer]{Hackmann:2014tga}
Hackmann, E.; L\"ammerzahl, C.; Obukhov, Y.N.; Puetzfeld, D.; Schaffer, I.
\newblock {Motion of spinning test bodies in Kerr spacetime}.
\newblock {\em Phys. Rev. D} {\bf 2014}, {\em 90},~064035, doi:10.1103/PhysRevD.90.064035.

\bibitem[Hendi \em{et~al.}(2020)Hendi, Tavakkoli, Panahiyan, Eslam~Panah, and
  Hackmann]{Hendi:2020yah}
Hendi, S.H.; Tavakkoli, A.M.; Panahiyan, S.; Eslam~Panah, B.; Hackmann, E.
\newblock {Simulation of geodesic trajectory of charged BTZ black holes in
  massive gravity}.
\newblock {\em Eur. Phys. J. C} {\bf 2020}, {\em 80},~524, doi:10.1140/epjc/s10052-020-8065-9.

\bibitem[Grunau and Kruse(2020)]{Grunau:2019bsd}
Grunau, S.; Kruse, M.
\newblock {Motion of charged particles around a scalarized black hole in
  Kaluza-Klein theory}.
\newblock {\em Phys. Rev. D} {\bf 2020}, {\em 101},~024051, doi:10.1103/PhysRevD.101.024051.

\bibitem[Soroushfar \em{et~al.}(2016)Soroushfar, Saffari, Kazempour, Grunau,
  and Kunz]{Soroushfar:2016esy}
Soroushfar, S.; Saffari, R.; Kazempour, S.; Grunau, S.; Kunz, J.
\newblock {Detailed study of geodesics in the Kerr-Newman-(A)dS spacetime and
  the rotating charged black hole spacetime in $f(R)$ gravity}.
\newblock {\em Phys. Rev. D} {\bf 2016}, {\em 94},~024052, doi:10.1103/PhysRevD.94.024052.

\bibitem[Hoseini \em{et~al.}(2016)Hoseini, Saffari, Soroushfar, Kunz, and
  Grunau]{Hoseini:2016nzw}
Hoseini, B.; Saffari, R.; Soroushfar, S.; Kunz, J.; Grunau, S.
\newblock {Analytic treatment of complete geodesics in a static cylindrically
  symmetric conformal spacetime}.
\newblock {\em Phys. Rev. D} {\bf 2016}, {\em 94},~044021, doi:10.1103/PhysRevD.94.044021.

\bibitem[Flathmann and Wassermann(2020)]{Flathmann:2019mlj}
Flathmann, K.; Wassermann, N.
\newblock {Geodesic equations for particles and light in the black spindle
  spacetime}.
\newblock {\em J. Math. Phys.} {\bf 2020}, {\em 61},~122504, doi:10.1063/5.0011432.

\bibitem[Chatterjee \em{et~al.}(2019)Chatterjee, Flathmann, Nandan, and
  Rudra]{Chatterjee:2019rym}
Chatterjee, A.K.; Flathmann, K.; Nandan, H.; Rudra, A.
\newblock {Analytic solutions of the geodesic equation for
  Reissner-Nordstr\"om\textendash{}(anti\textendash{})de Sitter black holes
  surrounded by different kinds of regular and exotic matter fields}.
\newblock {\em Phys. Rev. D} {\bf 2019}, {\em 100},~024044, doi:10.1103/PhysRevD.100.024044.

\bibitem[Kehagias and Sfetsos(2009)]{Kehagias:2009is}
Kehagias, A.; Sfetsos, K.
\newblock {The Black hole and FRW geometries of non-relativistic gravity}.
\newblock {\em Phys. Lett. B} {\bf 2009}, {\em 678},~123--126, doi:10.1016/j.physletb.2009.06.019.

\bibitem[Lü \em{et~al.}(2009)Lü, Mei, and Pope]{Lu:2009em}
Lü, H.; Mei, J.; Pope, C.N.
\newblock {Solutions to Horava Gravity}.
\newblock {\em Phys. Rev. Lett.} {\bf 2009}, {\em 103},~091301, doi:10.1103/PhysRevLett.103.091301.

\bibitem[Park(2009)]{Park:2009zra}
Park, M.
\newblock {The Black Hole and Cosmological Solutions in IR modified Horava
  Gravity}.
\newblock {\em JHEP} {\bf 2009}, {\em 09},~123, doi:10.1088/1126-6708/2009/09/123.

\bibitem[Boulware and Deser(1985)]{Boulware:1985wk}
Boulware, D.G.; Deser, S.
\newblock {String Generated Gravity Models}.
\newblock {\em Phys. Rev. Lett.} {\bf 1985}, {\em 55},~2656, doi:10.1103/PhysRevLett.55.2656.

\bibitem[Enolskii \em{et~al.}(2011)Enolskii, Hartmann, Kagramanova, Kunz,
  L\"ammerzahl, and Sirimachan]{Enolskii:2011rk}
Enolskii, V.; Hartmann, B.; Kagramanova, V.; Kunz, J.; L\"ammerzahl, C.;
  Sirimachan, P.
\newblock {Particle motion in Horava-Lifshitz black hole space-times}.
\newblock {\em Phys. Rev. D} {\bf 2011}, {\em 84},~084011, doi:10.1103/PhysRevD.84.084011.

\end{thebibliography}


%


\end{adjustwidth}
\end{document}